\documentclass[final,3p,11pt,authoryear]{elsarticle}

\makeatletter
\def\ps@pprintTitle{%
 \let\@oddhead\@empty
 \let\@evenhead\@empty
 \def\@oddfoot{\centerline{\thepage}}%
 \let\@evenfoot\@oddfoot}
\makeatother
\usepackage{color}
\usepackage[usenames,dvipsnames]{xcolor}
\usepackage{hyperref}

\usepackage{natbib}
\biboptions{numbers}
\usepackage[nodots]{numcompress}
\usepackage{graphicx}
\usepackage{caption}
\usepackage{subcaption}
\usepackage{amsfonts,amssymb,amsbsy,amsmath,mathtools}

\journal{International Journal of Solids and Structures}
\begin{document}

\begin{frontmatter}

\renewcommand{\thefootnote}{\fnsymbol{footnote}}
\title{\textbf{Hierarchy of beam models for lattice core sandwich structures\let\thefootnote\relax\footnote{{\color{Blue}\textbf{Recompiled, unedited accepted manuscript}}. \copyright 2020. Made available under \href{https://creativecommons.org/licenses/by-nc-nd/4.0/}{{\color{Blue}\textbf{\underline{CC-BY-NC-ND 4.0}}}}}}}


\author[add1]{Anssi T. Karttunen\corref{cor1}}
\cortext[cor1]{Corresponding author. anssi.karttunen@iki.fi. \textbf{Cite as}: \textit{Int. J. Solids Struct.} 2020;121:103423 \href{https://doi.org/10.1016/j.ijsolstr.2020.08.020}{{\color{OliveGreen}\textbf{\underline{doi link}}}}}
\address[add1]{Aalto University, Department of Mechanical Engineering, Espoo, Finland}
\author[add2]{J.N. Reddy}
\address[add2]{Texas A\&M University, Department of Mechanical Engineering, College Station, Texas, USA}




\begin{abstract}
A discrete-to-continuum transformation to model 2-D discrete lattices as energetically equivalent 1-D continuum beams is developed. The study is initiated in a classical setting but results in a non-classical two-scale micropolar beam model via a novel link within a unit cell between the second-order macrorotation-gradient and the micropolar antisymmetric shear deformation. The shear deformable micropolar beam is reduced to a couple-stress and two classical lattice beam models by successive approximations. The stiffness parameters for all models are given by the micropolar constitutive matrix. The four models are compared by studying stretching- and bending-dominated lattice core sandwich beams under various loads and boundary conditions. A classical 4th-order Timoshenko beam is an apt first choice for stretching-dominated beams, whereas the 6th-order micropolar model works for bending-dominated beams as well. The 6th-order couple-stress beam is often too stiff near point loads and boundaries. It is shown that the 1-D micropolar model leads to the exact 2-D lattice response in the absence of boundary effects even when the length of the 1-D beam (macrostructure) equals that of the 2-D unit cell (microstructure), that is, when $L=l$.
\end{abstract}

\end{frontmatter}


\section{Introduction}
Lattice materials are a class of lightweight structures that integrate concepts from materials science and structural mechanics. The dual nature of a lattice material is determined by a base material and a lattice structure driven by architected design \citep{greer2019}. 

Increasing interest in creating intricate lattice materials has triggered fresh thinking and new developments in the domain of manufacturing. Traditional manufacturing processes such as welding, investment casting or perforated metal sheet forming can be used to produce lattices visible to the naked eye \citep{wadley2006}. As a specific example, laser welding has been used to manufacture all-steel sandwich panels with lattice cores for large structures such as ship and bridge decks \citep{teasdale1988,roland1997,romanoff2007c,bright2007,nilsson2017,nilsson2020}. The production of man-made small-scale lattice materials from a variety of base materials at reasonable cost has been enabled in recent years by advances in additive manufacturing processes \citep{carter2016,valdevit2018,mines2019}. The different families of additive manufacturing are used to produce lattices in the areas of micro-electro-mechanical systems, metamaterials and biomechanics, to name a few \citep{phani2017}. 

From a structural point of view, lattice materials consist of periodic networks of rods, beams, plates and shells. For load-carrying applications, truss-like lattice constructions are usually considered. Such stretching-dominated configurations are designed to exhibit only axial compression and tension when subjected to external loads and are exceptionally stiff and strong, whereas bending-dominated lattices are a lot more flexible \citep{ashby2006}. Realistically speaking, any manufacturing process will introduce imperfections (e.g. member waviness) into a lattice material and, thus, an ideally stretching-dominated material may also exhibit bending in practice \citep{biagi2007,liu2017,pasini2019}. Lattices designed to be bending-dominated in their ideal state can be found in energy absorption applications, for example, see \citep{luc2015}. Lattice core sandwich structures are a trending research topic and, in addition to the above-referenced works, key engineering aspects of lattice core sandwich structures have also been discussed by \cite{birman2018} within context of recent applications of sandwich structures in general. In this paper, our focus is on the structural modeling of lattice core sandwich panels.

Typical stretching- and bending-dominated lattice materials can be modeled as discrete structures using beam and shell finite elements. Such models quickly become computationally expensive due to the large number of degrees of freedom involved. Thus, it is beneficial to first develop homogenized models so that the discrete lattices are modeled in an average sense as continuum materials. For such a continuum material description, it is necessary to pass information on both the translational and \textit{rotational} degrees of freedom from a discrete lattice model comprising beam and shell finite elements into the continuum. The tools necessary for this purpose are provided by couple-stress and micropolar continuum mechanics theories \citep{mindlin1962,toupin1962,mindlin1963,tiersten1974,yang2002,eringen1966,eringen2012}. 

Couple-stress and micropolar theories use macrorotation and microrotation, respectively, as rotational variables, whose first gradients, in turn, are used as strain measures to describe the bending of an infinitesimal material element, or unit cell. The sandwich beam bending example in Fig.~1 sheds light on the need for these theories in modeling lattice materials. The web-core beam is modeled as a 2-D FE beam frame using conventional beam elements. In the vicinity of point load $F$, the beam exhibits two types of bending; global bending with respect to the central axis and local bending with respect to the face (and web) axes. A classical 1-D beam model, which is a statically equivalent single-layer (ESL) description of the 2-D frame, can account only for the bending with respect to the central axis through the classical moment $M_{xx}$ and curvature $\kappa_{xx}$. In order to also take into account the local bending of the faces, a couple-stress moment $P_{xz}$ and curvature $\kappa_{xz}$, which depends on either the macrorotation $\omega_z$ (couple-stress theories) or the microrotation $\psi_z$ (micropolar theory), are needed. In conclusion, the need for non-classical theories is easy to understand on a conceptual level: a classical theory accounts for cell wall stretching, but a non-classical approach is needed for cell wall bending. Further study of the example would show that the main challenge in developing non-classical 1-D models for sandwich beams with lattice cores is the derivation of the bending and shear stiffnesses, i.e., the constitutive parameters.
\begin{figure}
\centering
\includegraphics[scale=0.48]{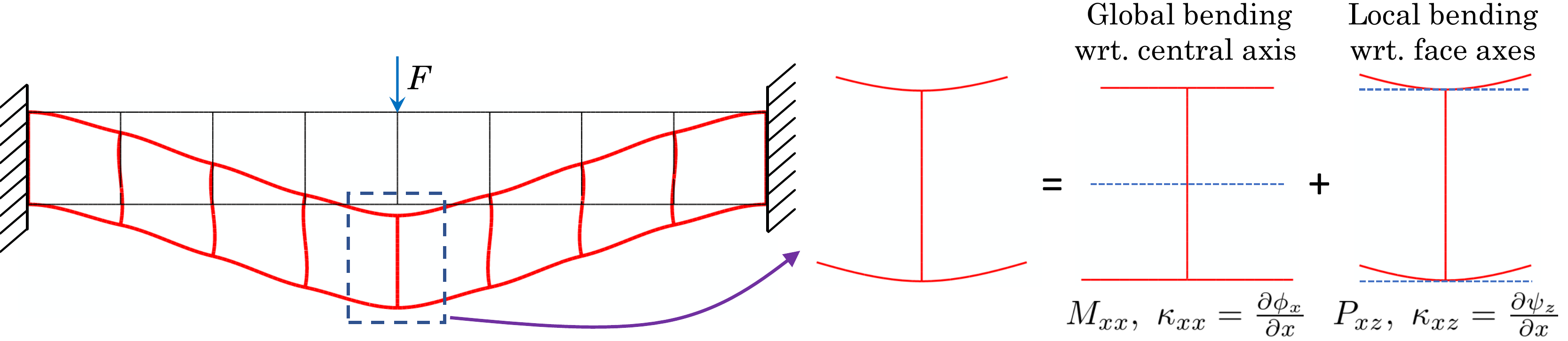}
\caption{Bending of a web-core beam modeled as a 2-D FE beam frame. In a 1-D beam, global bending is related to face sheet stretching/compression that determines the cross sectional rotation $\phi_x$ at the central axis. Non-classical 1-D beams can capture the local bending which depends on microrotation $\psi_z$ or macrorotation $\omega_z$.}
\end{figure}

The development of constitutive equations for structural models based on couple-stress and micropolar continuum theories has always been a major task. Motivated by high-rise buildings and other large gridworks, a few studies in the 1970s applied these theories to discrete lattice structures  \citep{bazant1972,sun1973,sun1975,kanatani1979}. In the 1980s, applications to space structures \citep{hefzy1986,noor1988} and also bones followed \citep{kim1987}. These early studies already raise the question: which approach should be used, couple-stress or micropolar? This question has lingered on to date; numerous studies have been conducted in recent years to determine the effective continuum parameters for lattice materials successfully according to either a couple-stress approach \citep{yoo2006,park2008,zhu2010,goda2015,rahali2016,ganghoffer2018,chen2019} or a micropolar approach \citep{kumar2004,chung2009,spadoni2012,dos2012,bacigalupo2014,chen2014,niu2016,reda2017,duan2018}. Several studies with particular emphasis on sandwich beams with lattice cores have been carried out as well in the past few years using either a couple-stress approach \citep{romanoff2014,penta2017,gesualdo2017,goncalves2018,goncalves2019} or a micropolar approach \citep{salehian2010,bacigalupo2019,karttunen2019a,nampally2019,chowdhury2019,liu2020}.  

In light of the preceding works on couple-stress and micropolar theories, the present study focuses on reviewing and comparing hierarchically derived 1-D micropolar, couple-stress and classical ESL beam models for sandwich beams with lattice cores. This is done to point out fundamental differences between the models and to facilitate not only the choice of a beam model for a particular application but the choice of a suitable continuum modeling framework for other lattice core sandwich structures such as plates and shells as well.

The rest of the paper is organized as follows. In Section 2, an energy-based, two-scale constitutive modeling method is used as the basis for deriving a 1-D micropolar ESL-FSDT (first-order shear deformation theory) beam model. In Section 3, the micropolar beam is reduced to a couple-stress model which is then further simplified to classical ESL-FSDT (Timoshenko) and Euler-Bernoulli ESL beam models by successive approximations. The stiffness parameters of an X-braced web-core for all beam models are given by the micropolar constitutive matrix. The core can be modified to be dominated by either stretching or bending in its ideal state. Initial imperfections that would add to the bending behavior of the lattice members are not considered in this study. In Section 4, general analytical solutions are developed for the four lattice beam models. The solutions are used in Section 5 to study lattice core sandwich beam bending problems. Both stretching- and bending-dominated lattice cores are considered. Finally, conclusions are drawn in Section 6. 

To elaborate further on the following sections, the two-scale constitutive modeling method based on a discrete-to-continuum transformation and the variants of the method have been used earlier for micropolar lattice core sandwich beams \citep{noor1980,noor1988,karttunen2019a,nampally2019}. However, as the central novel contribution, this method is reformulated in Section 2 to present the micropolar derivation as a spin-off of a more involved second-order macrorotation gradient beam model derivation. It is shown that there is a connection between the second-order macrorotation-gradient and the micropolar antisymmetric shear deformation within a general lattice material unit cell via a central difference approximation. This relation is presumed to have implications beyond 1-D beam models. In Section 3, the couple-stress beam is obtained as a simplification of the micropolar model, as discussed. However, in the current framework, the couple-stress model can also be derived as another independent spin-off of the approach that would lead to a 1-D second-order macrorotation gradient beam model. Finally, the connection between the second-order macrorotation-gradient and antisymmetric shear deformation also motivates, in part, the study of non-classical boundary effects in Sections 4 and 5. 
\section{Non-classical beam model for lattice core sandwich panels}
We formulate an equivalent single-layer, first-order shear deformation theory (ESL-FSDT) beam model for sandwich panels that have lattice cores. The derivation of the model is founded on a two-scale constitutive modeling approach. In this energy-based method, a discrete 2-D unit cell represents the \textit{microscale} and a 1-D continuum beam the \textit{macroscale}. Simply put, the unit cell characterizes the lattice material of which the beam is made of. The modeling approach can be briefly described by the following six steps which will be detailed further in subsections 2.1--2.6:
\begin{enumerate}
    \item[(1)] Displacements, rotations and strain measures based on classical elasticity are defined for a 1-D macroscale FSDT (Timoshenko) beam but no constitutive model is assumed a priori. 
    \item[(2)] A discrete 2-D microscale unit cell is attached to an arbitrary cross section of the 1-D macroscale beam. The \textit{discrete} unit cell corner displacements and rotations are presented in terms of the \textit{continuous} cross-sectional beam displacements and rotations via a Taylor series expansion. See Fig.~2a for an example of axial displacement expansion $U_x^i$ at node $i$.
    \item[(3)] A link is established within the unit cell between the second-order macrorotation-gradient based on \textit{classical} elasticity and the \textit{micropolar} antisymmetric shear deformation. This is done through a central difference approximation of the second-order macrorotation-gradient. Moreover, the first-order macrorotation-gradient is found to be approximately equal to the micropolar first-order microrotation-gradient. The resulting, effectively micropolar, beam strain measures are imposed on the gradients included in the Taylor series expansions derived in the previous step. See Fig.~2a for the kinematic imposition at node $i$. 
    \item[(4)] The kinematic imposition leads to a discrete-to-continuum transformation for the unit cell which in matrix form reads
    \begin{equation}
          \mathbf{d}=\mathbf{T}_u\mathbf{u}+\mathbf{T}_e\mathbf{e}
    \end{equation}
    where $\mathbf{d}$ is the generalized displacement vector of the discrete unit cell, $\mathbf{u}$ and $\mathbf{e}$ contain the continuous beam displacements and strains, respectively, and $\mathbf{T}_u$ and $\mathbf{T}_e$ are transformation matrices, the elements of which consist of ones, zeros and the unit cell length $l$ and height $h$. 
    By using the discrete-to-continuum transformation, the strain energy $W^d$ of the discrete microscale unit cell takes the form of the strain energy density $W^c_0$ of the unit cell in terms of the (micropolar) macroscale beam strains:
    \begin{equation}
\underbrace{W^{d}=\frac{1}{2}\mathbf{d}^{\textrm{T}}\mathbf{k}\mathbf{d}}_\text{discrete unit cell} \ \rightarrow \  \underbrace{W^{c}_0=\frac{1}{2}\mathbf{e}^{\textrm{T}}\mathbf{C}\mathbf{e}}_\text{continuum unit cell} \quad \textrm{with} \quad \mathbf{C}=\frac{1}{l}\mathbf{T}_e^{\textrm{T}}\mathbf{k}\mathbf{T}_e 
    \end{equation}
where $\mathbf{k}$ is the stiffness matrix of the unit cell and $\mathbf{C}$ is the constitutive matrix. Energy equivalence between the microscale unit cell and the macroscale beam is assumed by a hyperelastic constitutive relation through which micropolar beam stress resultants are defined as
    \begin{equation}
\mathbf{S}\equiv\frac{\partial W^c_0}{\partial \mathbf{e}}=\mathbf{C}\mathbf{e} 
\end{equation}
    \item[(5)] By further employing the cell-beam energy equivalence, the governing equations of a 1-D micropolar ESL-FSDT beam in terms of the stress resultants are derived using the principle minimum total potential energy.
    \item[(6)] Different discrete lattice materials may be modeled in detail by using classical beam finite elements in step (4) in order to obtain explicit constitutive matrices $\mathbf{C}$ for the presentation of the governing equations in terms of displacement and rotation variables. In this paper, we will consider an X-braced web-core unit cell modeled by Euler-Bernoulli beam (frame) finite elements. The core can be modified to be dominated by either stretching or bending. See Fig.~2b for a few examples of lattice material unit cells.
\end{enumerate}
\begin{figure}
\centering
\includegraphics[scale=1.2]{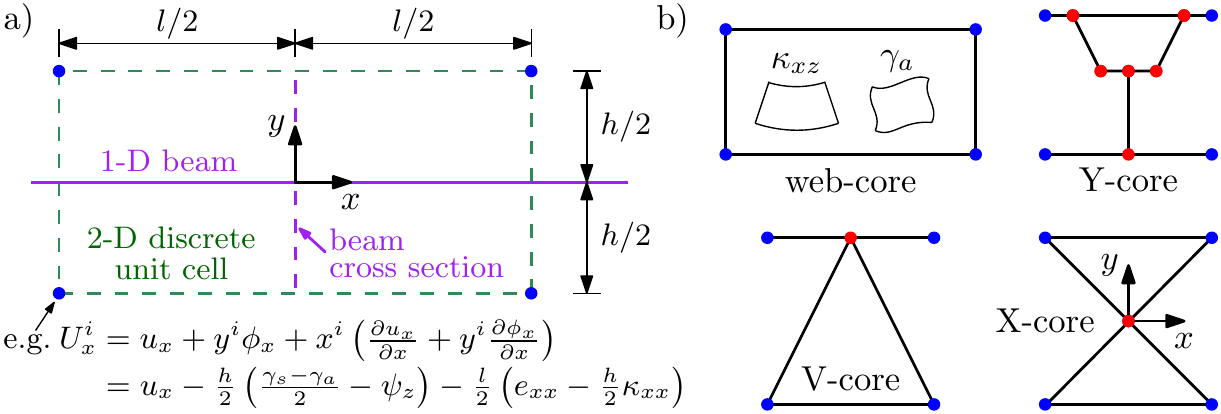}
\caption{a) General 2-D unit cell (microstructure) attached to an arbitrary cross section of a 1-D continuum beam (macrostructure). Discrete corner displacements of the 2-D unit cell are presented in terms of the cross sectional 1-D micropolar beam kinematics via a Taylor series expansion. b) Examples of unit cells for sandwich beams that conform to the general cell when all degrees of freedom in the red internal nodes have been condensed out. Non-classical beam strains, local curvature $\kappa_{xz}$ and antisymmetric shear strain $\gamma_a$, are presented within the web-core.}
\end{figure}
\subsection{Beam displacements, rotations and strains}
Displacements $U_x$ and $U_y$ of a conventional 1-D shear deformable (FSDT) beam can be expressed in terms of central axis kinematic variables ($u_x, u_y, \phi_x$) as
\begin{equation}
U_x(x,y)=u_x(x)+y\phi_x(x), \quad U_y(x,y)=u_y(x)
\end{equation}
where $u_x$ is the axial displacement, $\phi_x$ is the rotation of the cross section and $u_y$ is the transverse deflection. The macrorotation is given by
\begin{equation}
\Omega_z(x,y)=\frac{1}{2}\left(\frac{\partial U_y}{\partial x}-\frac{\partial U_x}{\partial y}\right)=\frac{1}{2}\left(\frac{\partial u_y}{\partial x}-\phi_x\right)=\omega_z(x)
\end{equation}
The classical nonzero infinitesimal strains of the beam are
\begin{align}
\varepsilon_{xx}&=\frac{\partial U_x}{\partial x}=\frac{\partial u_x}{\partial x}+y\frac{\partial \phi_x}{\partial x}=e_{xx}+y\kappa_{xx} \\
\gamma_{xy}&=\frac{\partial U_x}{\partial y}+\frac{\partial U_y}{\partial x}=\phi_x+\frac{\partial u_y}{\partial x}
\end{align}
where $e_{xx}$ and $\kappa_{xx}$ are the axial strain and global curvature of the beam, respectively, and $\gamma_{xy}$ is the transverse shear strain. 
\subsection{Displacement and rotation expansions}
Figure 2a shows a general, discrete rectangular unit cell attached to an arbitrary cross section of a beam. The unit cell corner displacements can be expressed in terms of the cross-sectional displacements $U_x$ and $U_y$, and the corner rotations by the aid of the macrorotation $\Omega_z$. With distance from an arbitrary beam cross section, first-order 1-D Taylor series expansions of the displacements (4) and a second-order expansion of the macrorotation (5) at the corner points $i$ located at $x^i=\pm l/2$ and $y^i=\pm h/2$ lead to
\begin{align}
U_x^i&=u_x+y^i\phi_x+x^i\left(\frac{\partial u_x}{\partial x}+y^i\frac{\partial \phi_x}{\partial x}\right) \\
U_y^i&=u_y+x^i\left(\frac{\partial u_y}{\partial x}\right) \\
\Omega_z^i&=\omega_z+x^i\left(\frac{\partial \omega_z}{\partial x}\right)+\frac{\left(x^i\right)^2}{2}\left(\frac{\partial^2 \omega_z}{\partial x^2}\right)
\end{align}
The above expansions (8)--(10) will lead to positive-definite constitutive matrices for different lattice materials that conform to the discrete unit cell displayed in Fig.~2a. The positive-definiteness guarantees that the materials are stable in the conventional sense (i.e., strain energy is positive for nonzero strains). Inclusion of any additional higher-order displacement or macrorotation-gradients into the expansions would violate the positive-definiteness and they are excluded from (8)--(10). We note that the four-node unit cell of Fig.~2 does not possess the degrees of freedom that would be necessary to represent the strain gradients brought about by any additional higher-order expansions. The positive-definiteness will be discussed further in Sections 2.3 and 2.6.

\subsection{Approximation of 2nd-order macrorotation-gradient and kinematic imposition}
The rotation expansion (10) includes third and second derivatives of $u_y$ and $\phi_x$, respectively. If these were to be included in the present developments, the total differential order of the bending part of the resulting beam model would be eight and the model would have non-conventional boundary conditions. To obtain a simpler model, we approximate the second-order macrorotation-gradient in the rotation expansion (10) by a three-point second-order central difference formula within the discrete unit cell so that
\begin{equation}
\frac{\partial^2 \omega_z}{\partial x^2}\approx\frac{\omega_z(x+l/2)-2\omega_z(x)+\omega_z(x-l/2)}{(l/2)^2}   
\end{equation}
Equation~(11) can be written as
\begin{equation}
\frac{\partial^2 \omega_z}{\partial x^2}\approx-\frac{4}{l^2}\gamma_a
\end{equation}
which includes the micropolar antisymmetric shear strain \citep{karttunen2019a}
\begin{equation}
\gamma_a=2(\omega_z-\psi_z)
\end{equation}
where the microrotation $\psi_z$ is now defined as the aritmethic mean of the macrorotations at the unit cell boundaries, that is,
\begin{equation}
\psi_z(x)\equiv\frac{\omega_z(x+l/2)+\omega_z(x-l/2)}{2}    \end{equation}
Furthermore, Eqs.~(12) and (13) give
\begin{equation}
\frac{\partial \psi_z}{\partial x}\approx\frac{\partial \omega_z}{\partial x}+\frac{l^2}{8}\left(\frac{\partial^3 \omega_z}{\partial x^3}\right)   
\end{equation}
As discussed at the end of the previous section, the inclusion of the third-order macrorotation-gradient into Eqs.~(8)--(10) would violate the positive-definiteness of the constitutive matrix that will be derived and discussed in the following sections. Therefore, in order to obtain a physically plausible beam model, we assume
\begin{equation}
\frac{\partial \psi_z}{\partial x}\approx\frac{\partial \omega_z}{\partial x}  
\end{equation}
For a graphical representation of the macrorotation-gradient, see \citep{mindlin1963}. 

We define the following strain notations
\begin{equation}
\gamma_{s}\equiv\gamma_{xy} 
\end{equation} 
\begin{equation}
\kappa_{xz}\equiv\frac{\partial \psi_z}{\partial x}
\end{equation}
where $\gamma_s$ is the symmetric shear strain that accompanies the antisymmetric shear $\gamma_a$ and $\kappa_{xz}$ is the local curvature of the beam in contrast to its global curvature $\kappa_{xx}$ in Eq.~(6). By imposing the strains $(e_{xx},\kappa_{xx},\kappa_{xz},\gamma_s,\gamma_a)$ from Eqs.~(6), (13) and (18) on the expansions (8)--(10), we obtain
\begin{align}
U_x^i&=u_x+y^i\left(\frac{\gamma_s-\gamma_a}{2}-\psi_z\right)+x^i\left(e_{xx}+y^i\kappa_{xx}\right) \\
U_y^i&=u_y+x^i\left(\frac{\gamma_s+\gamma_a}{2}+\psi_z\right) \\
\Omega_z^i&=\psi_z+x^i(\kappa_{xz})
\end{align}
The microrotation $\psi_z$ practically replaces the macrorotation $\omega_z$ through Eqs.~(12) and (13) and the classical macrorotation and its gradients are no longer explicitly present in the expansions (19)--(21), which are exactly the same as the ones developed in the context of micropolar elasticity \citep{karttunen2019a}. The ingenuity of the micropolar approach lies in the fact that higher-order displacement-based derivatives included in the macrorotation-gradients are essentially replaced by the microrotation and its first (low-order) derivative which simplifies the resulting beam model. It is noteworthy that, in general, the microrotation in the micropolar theory is considered to be independent of the displacements \citep{eringen2012}. However, definition (14) connects the micropolar microrotation and the classical macrorotation within the unit cell.

The above treatment also effectively changes the definition of the relative strains from classical (cl) to micropolar (mp) so that
\begin{align}
\varepsilon_{xy}^{cl}&=\frac{\partial U_y}{\partial x}-\Omega_z=\frac{\partial u_y}{\partial x}-\omega_z \quad \: \rightarrow \quad \varepsilon_{xy}^{mp}=\frac{\partial u_y}{\partial x}-\psi_z, \\
\varepsilon_{yx}^{cl}&=\frac{\partial U_x}{\partial y}+\Omega_z=\phi_x+\omega_z \qquad \rightarrow \quad \varepsilon_{yx}^{mp}=\phi_x+\psi_z 
\end{align}
but it is important to note that in general
\begin{equation}
\psi_z\neq\omega_z    
\end{equation}
The symmetric and antisymmetric shear strains are defined as
\begin{align}
\gamma_s&=\varepsilon_{xy}^{mp}+\varepsilon_{yx}^{mp}=\frac{\partial u_y}{\partial x}+\phi_x \\    \gamma_a&=\varepsilon_{xy}^{mp}-\varepsilon_{yx}^{mp}=\frac{\partial u_y}{\partial x}-\phi_x-2\psi_z=2(\omega_z-\psi_z)
\end{align}
respectively. The antisymmetric shear may also be referred to as the \textit{rotational deformation} \citep{debellis2011} and it describes a similar strain as the second-order macrorotation-gradient. In Section 3, the micropolar beam theory will be reduced to a couple-stress beam theory. In general, the micropolar theory of elasticity reduces to the couple-stress theory when the microrotation and macrorotation coincide, that is, $\psi_z=\omega_z$ \citep{sadd2014}.
\subsection{Two-scale constitutive modeling}
The discrete-to-continuum transformation (19)--(21) can be written in matrix form as [Eq.~(1) is repeated here for the sake of convenience]
\begin{equation}
\mathbf{d}=\mathbf{T}_u\mathbf{u}+\mathbf{T}_e\mathbf{e}
\end{equation}
where the generalized discrete displacement vector $\mathbf{d}$ consists of the degrees of freedom at the unit cell corner nodes, and the vectors for the continuous variables read
\begin{align}
\mathbf{u}&=\left\{u_x \ \ u_y \ \ \phi_x \ \ \psi_z\right\}^{\textrm{T}} , \\
\mathbf{e}&=\{e_{xx} \quad \kappa_{xx} \quad \kappa_{xz} \quad \gamma_s \quad \gamma_a\}^{\textnormal{T}} 
\end{align}
The displacement and strain transformation matrices are
\begin{equation}
\mathbf{T}^{\phantom{ }}_u=
\left[
\begin{array}{cccccccccccc}
 1 & 0 & 0 & 1 & 0 & 0 & 1 & 0 & 0 & 1 & 0 & 0 \\
 0 & 1 & 0 & 0 & 1 & 0 & 0 & 1 & 0 & 0 & 1 & 0 \\
 0 & 0 & 0 & 0 & 0 & 0 & 0 & 0 & 0 & 0 & 0 & 0 \\
 \frac{h}{2} & -\frac{l}{2} & 1 & \frac{h}{2} & \frac{l}{2} & 1 & -\frac{h}{2} & \frac{l}{2} & 1 & -\frac{h}{2} & -\frac{l}{2} & 1
   \\
\end{array}
\right]^{\textrm{T}}
\end{equation}
and
\begin{equation}
\mathbf{T}^{\phantom{ }}_e=
\left[
\begin{array}{cccccccccccc}
 -\frac{l}{2} & 0 & 0 & \frac{l}{2} & 0 & 0 & \frac{l}{2} & 0 & 0 & -\frac{l}{2} & 0 & 0 \\
 \frac{h l}{4} & 0 & 0 & -\frac{h l}{4} & 0 & 0 & \frac{h l}{4} & 0 & 0 & -\frac{h l}{4} & 0 & 0 \\
 0 & 0 & -\frac{l}{2} & 0 & 0 & \frac{l}{2} & 0 & 0 & \frac{l}{2} & 0 & 0 & -\frac{l}{2} \\
 -\frac{h}{4} & -\frac{l}{4} & 0 & -\frac{h}{4} & \frac{l}{4} & 0 & \frac{h}{4} & \frac{l}{4} & 0 & \frac{h}{4} & -\frac{l}{4} & 0
   \\
 \frac{h}{4} & -\frac{l}{4} & 0 & \frac{h}{4} & \frac{l}{4} & 0 & -\frac{h}{4} & \frac{l}{4} & 0 & -\frac{h}{4} & -\frac{l}{4} & 0
   \\
\end{array}
\right]^{\textrm{T}}
\end{equation}
respectively. The strain energy of a discrete (typically FE-based) unit cell can be written as
\begin{equation}
W^d=\frac{1}{2}\mathbf{d}^{\textrm{T}}\mathbf{k}\mathbf{d}
\end{equation}
where $\mathbf{k}$ is the global stiffness matrix of the unit cell and corresponds to the degrees of freedom in the unit cell corners after static condensation has been carried out in the internal nodes (if they exist). By applying the transformation (27) to Eq.~(32), the strain energy becomes
\begin{equation}
W^c=\frac{1}{2}\mathbf{e}^{\textrm{T}}\mathbf{T}_e^{\textrm{T}}\mathbf{k}\mathbf{T}^{\phantom{ }}_e\mathbf{e}
\end{equation}
The displacements $\mathbf{u}$ in the transformation (27) do not contribute to the unit cell strain energy (33) as they relate to rigid body modes. Furthermore, this can be verified in the case of specific lattice cores such as the one in Section 2.6 and those studied by \cite{nampally2019}. We define the \textit{linear density} of the unit cell strain energy as
\begin{equation}
W_0^c\equiv\frac{W^c}{l}=\frac{1}{2}\mathbf{e}^{\textrm{T}}\mathbf{C}\mathbf{e}
\end{equation}
where the constitutive matrix is given by
\begin{equation}
\mathbf{C}=\frac{1}{l}\mathbf{T}_e^{\textrm{T}}\mathbf{k}\mathbf{T}^{\phantom{ }}_e
\end{equation}
The unit cell represents a \textit{lattice material} of which the micropolar beam is made of. Therefore, in analogy with any hyperelastic material, we write for the micropolar beam continuum
\begin{equation}
\mathbf{S}\equiv\frac{\partial W_0^c}{\partial \mathbf{e}}=\mathbf{C}\mathbf{e}
\end{equation}
where the stress resultant vector of the micropolar beam is
\begin{equation}
\mathbf{S}=\{N_{xx} \quad M_{xx} \quad P_{xz} \quad Q_s \quad Q_a\}^{\textnormal{T}}    
\end{equation}
where, for sandwich beams, $N_{xx}$ is the axial force, $M_{xx}$ is the global bending moment related to bending with respect to the central axis of a sandwich beam, whereas $P_{xz}$ is the local bending moment associated with the bending of the sandwich face sheets with respect to their own central axes. The symmetric and antisymmetric shear forces are denoted by $Q_s$ and $Q_a$, respectively.
\subsection{Governing equations}
We derive the equilibrium equations for the 1-D micropolar ESL-FSDT beam model by using the principle of total potential energy. The constitutive relation (36) established an energy equivalence between the macrostructure (beam) and the microstructure (unit cell). Now, in a similar fashion, using the unit cell strain energy density $W_0^c$, we write the strain energy for a micropolar beam of length $L$ as
\begin{equation}
U=\int_{0}^{L} W_0^c\ dx=\frac{1}{2}\int_{0}^{L} \mathbf{e}^{\textrm{T}}\mathbf{C}\mathbf{e}\ dx
\end{equation}
The potential energy contribution due to a distributed transverse load $q(x)$ and a distributed moment $m(x)$ exerted on the beam is
\begin{equation}
V=-\int_{0}^{L}(qu_y+m\psi_z)\ dx.
\end{equation}
By using Eqs.~(38) and (39) and the principle of total potential energy \citep{reddy2018}, we have
\begin{equation}
\delta(U+V)=0,
\end{equation}
which we can write in the form
\begin{equation}
\int_{0}^{L}\delta\mathbf{e}^{\textrm{T}}\mathbf{S}-q\delta u_y-m\delta\psi_z\ dx=0
\end{equation}
where $\mathbf{S}=\mathbf{Ce}$. In explicit form, we have
\begin{equation}
\int_{0}^{L}\left(N_{xx}\delta e_{xx}+M_{xx}\delta \kappa_{xx}+P_{xz}\delta \kappa_{xz}+Q_s\delta\gamma_{s}+Q_a\delta\gamma_{a}-q\delta u_y-m\delta\psi_z\right)dx=0
\end{equation}
or equivalently
\begin{equation}
\begin{aligned}
\int_{0}^{L}&\bigg[N_{xx}\frac{\partial\delta u_x}{\partial x}+M_{xx}\frac{\partial\delta \phi_x}{\partial x}+P_{xz}\frac{\partial\delta \psi_z}{\partial x}+Q_s\left(\frac{\partial\delta u_y}{\partial x}+\delta\phi_x\right)\\&+Q_a\left(\frac{\partial\delta u_y}{\partial x}-\delta\phi_x-2\delta\psi_z\right)-q\delta u_y-m\delta\psi_z\bigg]dx=0
\end{aligned}
\end{equation}
We arrive at the following equilibrium equations (Euler--Lagrange equations) of the 1-D micropolar beam essentially by applying integration by parts in Eq.~(43):
\begin{align}
\frac{\partial N_{xx}}{\partial x}&=0 \\ 
\frac{\partial }{\partial x}(Q_s+Q_a)&=-q \\
\frac{\partial M_{xx}}{\partial x}-Q_s+Q_a&=0 \\
\frac{\partial P_{xz}}{\partial x}+2Q_a&=-m
\end{align}
As for the boundary conditions at the beam ends, one element in each of the following four duality pairs should be specified at $x=0$ and $x=L$
\begin{align}
N_x \quad &\textrm{or} \quad u_x \\  Q_s+Q_a \quad &\textrm{or} \quad u_y \\ M_{xx} \quad &\textrm{or} \quad \phi_x \\  P_{xz} \quad &\textrm{or} \quad \psi_z
\end{align}
Alternative definitions for the shear forces are \citep{mindlin1963,karttunen2019a}
\begin{equation}
Q_{xy}=Q_s+Q_a \quad \textrm{and} \quad Q_{yx}=Q_s-Q_a 
\end{equation}
Figure~3 shows the stress resultants acting on the beam and the shear force definitions.
\begin{figure}[h!]
\centering
\includegraphics[scale=0.75]{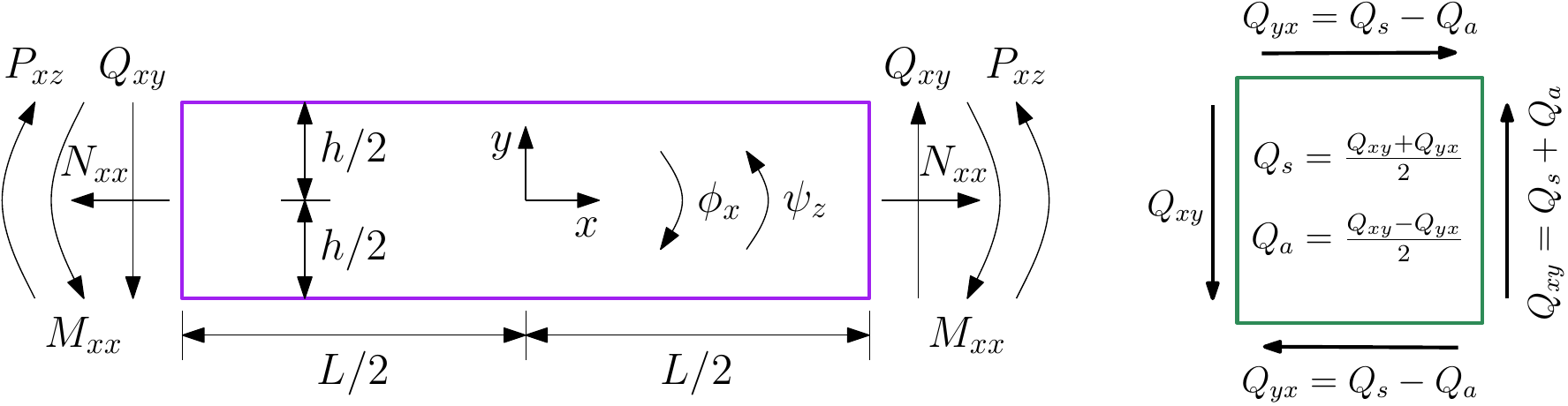}
\caption{1-D micropolar ESL-FSDT beam. The positive directions of the stress resultants and rotations are shown as well as the split of the shear forces into symmetric and antisymmetric parts \citep{mindlin1963}.}
\end{figure}
\subsection{Constitutive matrix for X-braced web-core}
The X-braced web-core unit cell displayed in Fig.~4 is modeled by eight nodally-exact linearly elastic isotropic, homogeneous Euler--Bernoulli beam finite elements with rectangular cross sections. The unit cell has similar top and bottom faces with axial stiffness $EA_f$ and bending stiffness $EI_f$. The diagonal elements have axial and bending stiffnesses $EA_d$ and $EI_d$, respectively. Only half of each web stiffness parameter is accounted for due to symmetry between neighboring unit cells so that we have $EA_w/2$, $EI_w/2$ and $k_\theta/2$ for the axial, bending and rotational joint stiffnesses, respectively. While the faces and diagonals are modeled using conventional Euler--Bernoulli beam elements, the webs are modeled using special-purpose elements with rotational springs at both ends to account for the flexibility of the joints \citep{monforton1963,chen2005}. The flexibility can be attributed, for example, to the fact that the joints between the webs and faces are laser-welded so that the weld is thinner than the web \citep{romanoff2007c}. Static condensation is applied at the internal center-point node so that the global stiffness matrix $\mathbf{k}$ conforms to the discrete unit cell of Fig.~2a and the discrete-to-continuum transformation (27). 
\begin{figure}[hb]
\centering
\includegraphics[scale=1.2]{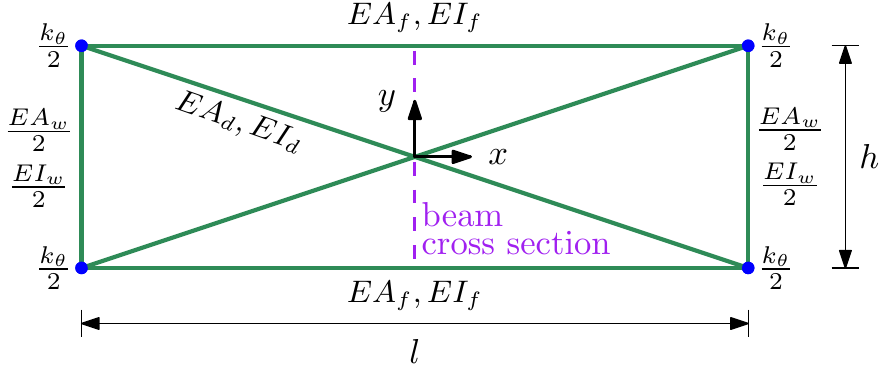}
\caption{Arbitrary cross section of the micropolar beam of length $L$ possessing X-braced web-core microstructure of length $l$. Static condensation has been applied at the center-point node.}
\end{figure}

The strain energy of the finite element unit cell can be written in the form of Eq.~(32) after which we apply the transformation (27) to obtain the constitutive matrix (35)
\begin{equation}
\mathbf{C}=\mathbf{C}^\textrm{web}+\mathbf{C}^\textrm{X}
\end{equation}
which is symmetric and where the web-core parameters are included in
\begin{equation}
\mathbf{C}^\textrm{web}=
\left[
\begin{array}{ccccc}
 2 EA_f & 0 & 0 & 0 & 0 \\
 0 & \frac{EA_f h^2}{2}+\Theta  & \Theta  & 0 & 0 \\
 0 & \Theta  & 2 EI_f+\Theta  & 0 & 0 \\
 0 & 0 & 0 & \frac{6 EI_f+\Theta }{l^2} & \frac{6 EI_f-\Theta
   }{l^2} \\
 0 & 0 & 0 & \frac{6 EI_f-\Theta }{l^2} & \frac{6 EI_f+\Theta
   }{l^2} \\
\end{array}
\right]    
\end{equation}
with 
\begin{equation}
\Theta=\frac{3EI_w k_\theta l}{6EI_w+k_\theta h}
\end{equation}
and the contributions from the X-braces are
\begin{equation}
\mathbf{C}^\textrm{X}=
\left[
\begin{array}{ccccc}
 \frac{16 \zeta_3 h^2 l^2}{\zeta_1^2}+\frac{2EA_d l^3}{\zeta_1
   \sqrt{\zeta_1}} & 0 & 0 & 0 & 0 \\
 0 & 2\zeta_2\sqrt{\zeta_1}lh^2 & \zeta_2\sqrt{\zeta_1}lh^2 & 0
   & 0 \\
 0 &  \zeta_2\sqrt{\zeta_1}lh^2  & \frac{\zeta_3 l^2}{3} 
   \left(1+\frac{\zeta_2 \zeta_1 h^2}{4 EI_d}\right) & 0 & 0 \\
 0 & 0 & 0 & \frac{\zeta_3 \left(h^2-l^2\right)^2}{\zeta_1^2}+\frac{2 EA_d lh^2
   }{\zeta_1 \sqrt{\zeta_1}} & \frac{\zeta_3
   \left(l^2-h^2\right)}{\zeta_1} \\
 0 & 0 & 0 & \frac{\zeta_3\left(l^2-h^2\right)}{\zeta_1} & \zeta_3 \\
\end{array}
\right]
\end{equation}
where
\begin{align}
\zeta_1&=h^2+l^2 \\
\zeta_2&=\frac{12EA_d EI_d}{48 EI_d l^2+EA_d h^2 \zeta_1} \\
\zeta_3&=\frac{6EI_d}{l\sqrt{\zeta_1}}
\end{align}
The constitutive matrix (53) forms the constitutive relations (36) together with the strain and stress resultant vectors given by Eqs.~(29) and (37), respectively. It is straightforward to verify by using the transformation (27) with and without the displacement part $\mathbf{T}_u\mathbf{u}$ that the displacements do not contribute to the strain energy density of the unit cell (34). Furthermore, the leading principal minors of the constitutive matrix (53) are positive and nonzero and, thus, the matrix is positive-definite and represents a stable lattice material in the conventional sense ($U>0$ for nonzero $\mathbf{e}$). In fact, the leading principal minors do not include any negative terms in symbolic form. For details on the positive-definiteness of web-core lattice material, in particular, see \citep{karttunen2019a}. If the third-order macrorotation-gradient in Eq.~(15) had been included in the expansions (19)--(21), the resulting constitutive matrix would have contained one leading principal minor equal to zero.

If we now consider only the web-core, Eq.~(54) shows that the global bending moment $M_{xx}$ is associated with $EA_fh^2/2$ under $C^{\textrm{web}}_{22}$ which is the usual global bending stiffness generated by the sandwich effect \citep{allen1969,zenkert1997}. The local bending moment $P_{xz}$, on the other hand, contains the bending stiffness $2EI_f$ of the two face sheets with respect to their own central axes. The global and local moments are coupled by $\Theta$ given in Eq.~(54). 

The diagonal X-braces increase the shear stiffness of the unit cell considerably. For $h=l$, the symmetric shear stiffness parameter in Eq.~(56) simplifies to
\begin{equation}
C^\textrm{X}_{44}=\frac{EA_d}{\sqrt{2}}    
\end{equation}
which means that the diagonals carry the shear with their axial stiffness and the unit cell is \textit{stretching-dominated}. In addition, the coupling between the symmetric and antisymmetric shear in Eq.~(56) vanishes for $h=l$. If the X-braces were to be removed, the unit cell would become \textit{bending-dominated}. This can be seen from the symmetric shear stiffness $C^\textrm{web}_{44}$ in Eq.~(54), which does not include any axial stiffness terms that would be large in comparison to bending stiffnesses $EI_f$ and $EI_w$. 

If we were to consider, for example, a Y-core that does not possess symmetry with respect to the beam central axis, the resulting symmetric constitutive matrix would also include a nonzero membrane-bending coupling term $C_{12}$, see \citep{nampally2019}. Finally, we note that the shear stiffnesses $C_{44}$, $C_{45}$ and $C_{55}$ account for the shear behavior of the micropolar ESL-FSDT (ESL-Timoshenko) beam model as well as possible through the beam-cell energy equivalence scheme and, thus, the model does not employ any extrinsic micropolar shear correction factors. 

\section{A hierarchy of sandwich beam models with lattice cores}
The 1-D micropolar ESL-FSDT beam model can be systematically reduced first to a couple-stress type of a model that can be then further simplified to a classical ESL-FSDT beam which in turn may be boiled down to an Euler-Bernoulli ESL beam. The simplification procedure amounts to static condensation of the constitutive equation (36) with the strains as the degrees of freedom. This same approach was used by \cite{noor1980} and \cite{noor1988}. From a physical point of view, the method first assumes that an internal force (beam stress resultant) is equal to zero, which implies that the associated deformations are allowed to occur freely. This in turn results in a situation where certain strains can be expressed in terms of other strain components so that we effectively end up having fewer strain components overall and, thus, a route to a simplified beam theory. The axial behavior is the same for each beam model and, thus, we consider only the bending parts of the beams. The differences and areas of application of the four different ESL beam models will be studied in Section 5 by numerical examples.
\subsection{Couple-stress ESL-FSDT beam model}
The micropolar constitutive equations for the bending part of the X-braced web-core beam in matrix form are 
\begin{equation}
\begin{Bmatrix}
M_{xx} \\
P_{xz} \\
Q_s \\
Q_a
\end{Bmatrix}
=
\begin{bmatrix}
C_{22} & C_{23} & 0 & 0  \\
C_{23} & C_{33} & 0 &0 \\
 0 & 0 & C_{44} & C_{45} \\
 0 & 0 & C_{45} & C_{55}
\end{bmatrix}
\begin{Bmatrix}
\kappa_{xx} \\
\kappa_{xz} \\
\gamma_s \\
\gamma_a 
\end{Bmatrix}
\end{equation}
We assume that the internal antisymmetric shear force is zero ($Q_a=0$) along the beam under any external loads. Then the last row in Eq.~(61) gives
\begin{equation}
\gamma_a=-\frac{C_{45}}{C_{55}}\gamma_s
\end{equation}
which on the second to last row yields
\begin{equation}
Q_s=\left(C_{44}-\frac{C_{45}^2}{C_{55}}\right)\gamma_s    
\end{equation}
The constitutive relations in Eq.~(61) reduce to
\begin{equation}
\begin{Bmatrix}
M_{xx} \\
P_{xz} \\
Q_s
\end{Bmatrix}
=
\begin{bmatrix}
 C_{22} & C_{23} & 0  \\
 C_{23} & C_{33} & 0 \\
 0 & 0 & C_{44}-\frac{C_{45}^2}{C_{55}}  \\

\end{bmatrix}
\begin{Bmatrix}
\kappa_{xx} \\
\kappa_{xz} \\
\gamma_s
\end{Bmatrix}
\end{equation}
which is still of the form $\mathbf{S}=\mathbf{C}\mathbf{e}$. The assumption $Q_a=0$ (under any load) necessitates the rederivation of the equilibrium equations. We take the strain energy density to be given as in Eq.~(34). In addition, we redefine Eq.~(18) as
\begin{equation}
\kappa_{xz}\equiv\frac{\partial \omega_z}{\partial x}   
\end{equation}
by using the unit cell approximation (16). Without a contribution from a distributed moment, the variational statement is then [cf.\ Eq.~(43)]
\begin{equation}
\int_{0}^{L}M_{xx}\frac{\partial\delta \phi_x}{\partial x}+\frac{P_{xz}}{2}\left(\frac{\partial^2\delta u_y}{\partial x^2}-\frac{\partial\delta \phi_x}{\partial x}\right)+Q_s\left(\frac{\partial\delta u_y}{\partial x}+\delta\phi_x\right)-q\delta u_y\ dx=0
\end{equation}
which results in the following equilibrium equations for a 1-D couple-stress beam:
\begin{align} 
\frac{\partial Q_s}{\partial x}-\frac{1}{2}\frac{\partial^2 P_{xz}}{\partial x^2}&=-q \\
\frac{\partial M_{xx}}{\partial x}-\frac{1}{2}\frac{\partial P_{xz}}{\partial x}-Q_s&=0
\end{align}
As for the boundary conditions at the beam ends, one element in each of the following three duality pairs should be specified 
\begin{align}
Q_s-\frac{1}{2}\frac{\partial P_{xz}}{\partial x} \quad &\textrm{or} \quad u_y \\ M_{xx} \quad &\textrm{or} \quad \phi_x \\  P_{xz} \quad &\textrm{or} \quad \omega_z
\end{align}
The derived beam model is similar to beam models based on the modified couple-stress theory \citep{reddy2011,asghari2011}. However, in the present case the constitutive relations are founded on a specific, well-defined microstructure (Fig.~4) and the model does not include a microstructural length-scale parameter that needs to be determined. In fact, complex lattice materials may involve numerous length-scale and other material parameters that describe the properties of the associated discrete unit cell. 

Expansions (8)--(10) can also be used to obtain exactly the same couple-stress beam model as derived above. In this approach, the macrorotation (5) and the strains (6) and (7) are imposed on the expansions (8)--(10). Then the constitutive relations are derived in the same manner as in the micropolar case. To obtain the couple-stress model, it is assumed that the stress resultant, say, $R_{xz}$ energetically conjugate to the second-order macrorotation-gradient, is zero $(R_{xz}=0)$. Constitutive equations (64) are obtained and the variational statement (66) follows. Although used as the starting point, the derivation of the second-order macrorotation-gradient beam model was not completed in this study nor does there exist such a beam model for lattice core beams in the literature. Nevertheless, a comprehensive treatment of a second-order macrorotation-gradient continuum mechanics theory has been given by \cite{shaat2016}. 
\subsection{Classical ESL-FSDT beam model}
We consider the matrix equation (64). Now the internal local bending moment is assumed zero $(P_{xz}=0)$ along the beam under any external loads. This assumption leads to
\begin{equation}
\kappa_{xz}=-\frac{C_{23}}{C_{33}}\kappa_{xx}    
\end{equation}
The resulting constitutive equations in matrix form read
\begin{equation}
\begin{Bmatrix}
M_{xx} \\
Q_s
\end{Bmatrix}
=
\begin{bmatrix}
C_{22}-\frac{C_{23}^2}{C_{33}} & 0  \\
0 & C_{44}-\frac{C_{45}^2}{C_{55}} 
\end{bmatrix}
\begin{Bmatrix}
\kappa_{xx} \\
\gamma_s
\end{Bmatrix}
\end{equation}
The variational statement is [cf.\ Eq.~(66)]
\begin{equation}
\int_{0}^{L}M_{xx}\frac{\partial\delta \phi_x}{\partial x}+Q_s\left(\frac{\partial\delta u_y}{\partial x}+\delta\phi_x\right)-q\delta u_y\ dx=0
\end{equation}
which results in the following equilibrium equations for a 1-D classical ESL-FSDT beam:
\begin{align}
\frac{\partial Q_s}{\partial x}&=-q \\
\frac{\partial M_{xx}}{\partial x}&=Q_s
\end{align}
The boundary conditions at the beam ends are
\begin{align}
Q_s \quad &\textrm{or} \quad u_y \\ M_{xx} \quad &\textrm{or} \quad \phi_x
\end{align}
Unlike the 6th-order micropolar and couple-stress beam models, the classical ESL-FSDT has a 4th-order bending part and only two boundary conditions are defined at each end of the beam. The novel feature for the classical model here is the constitutive matrix defined in Eq.~(73). 
\subsection{Classical Euler--Bernoulli ESL beam model}
For an Euler-Bernoulli beam model the constitutive equations (73) simplify to the global bending moment equation
\begin{equation}
M_{xx}=\left(C_{22}-\frac{C_{23}^2}{C_{33}}\right)\kappa_{xx} 
\end{equation}
and $\phi_x=-\frac{\partial u_y}{\partial x}$ so that $\kappa_{xx}=-\frac{\partial^2 u_y}{\partial x^2}$ and the equilibrium equation is
\begin{equation}
\frac{\partial^2 M_{xx}}{\partial x^2}=-q
\end{equation}
For the boundary conditions at the beam ends, we have
\begin{align}
\frac{\partial M_{xx}}{\partial x}\ \left(\equiv Q_{xy}\right) \quad &\textrm{or} \quad u_y \\ M_{xx} \quad &\textrm{or} \quad \frac{\partial u_y}{\partial x}
\end{align}
In conclusion, a hierarchy of lattice core sandwich beam models was established above by deriving three simplified models from the micropolar model by successive approximations. The micropolar constitutive matrix (53) includes the stiffness parameters for all models. 
\section{General analytical solutions and rotation relations}
General analytical solutions are presented here for the 1-D micropolar, couple-stress and classical beam models. Using the micropolar beam solution, the relations between the microrotation and macrorotation established in Section 2.3 for the microscale unit cell are studied.
\subsection{Classical Euler-Bernoulli ESL beam model}
The general solution to Eq.~(80) for a uniformly distributed load $q=q_0$ is
\begin{equation}
u_y=a_1+a_2 x+\frac{a_3 x^2}{2}+\frac{a_4 x^3}{6}+q_{uy}^{(1)} \end{equation}
where 
\begin{equation}
q_{uy}^{(1)}=\frac{C_{33}q_0x^4}{24(C_{22}C_{33}-C_{23}^2)}
\end{equation}
and the constants $a_i$ ($i=1,2,3,4$) are determined using boundary conditions (81) and (82).
\subsection{Classical ESL-FSDT beam model}
The general solution to Eqs.~(75) and (76) for a uniformly distributed load $q=q_0$ is
\begin{align}
u_y&=b_1+b_2x+\frac{b_3x^2}{2}+\frac{b_4 x^3}{6}+q_{uy}^{(1)} \\
\phi_x&=-b_2-b_3 x-b_4 \left(\frac{x^2}{2}+\delta_1\right)-\frac{\partial q_{uy}^{(1)}}{\partial x}-\delta_1\frac{\partial^3 q_{uy}^{(1)}}{\partial x^3}
\end{align}
where
\begin{equation}
\delta_1=\frac{C_{55}(C_{23}^2-C_{22}C_{33})}{C_{33}(C_{45}^2-C_{44}C_{55})}
\end{equation}
and the constants $b_i$ ($i=1,2,3,4$) are determined by the aid of boundary conditions (77) and (78).
\subsection{Couple-stress ESL-FSDT beam model}
The general solution to Eqs.~(67) and (68) for a uniformly distributed load $q=q_0$ is
\begin{align}
u_y&=c_1+c_2x+c_3x^2+c_4x^3+c_5e^{\alpha_1x}+c_6e^{-\alpha_1 x}+q_{uy}^{(2)} \\
\phi_x&=-c_2-2c_3x-3c_4\left(x^2+\alpha_3\right)-\alpha_1\alpha_2\left(c_5e^{\alpha_1x}-c_6e^{-\alpha_1x}\right)-\frac{\partial q_{uy}^{(2)}}{\partial x}-\frac{\alpha_3}{2}\frac{\partial^3 q_{uy}^{(2)}}{\partial x^3}
\end{align}
where
\begin{align}
\alpha_1&=\frac{2\sqrt{C_{22}-2C_{23}+C_{33}}\sqrt{C_{44}C_{55}-C_{45}^2}}{\sqrt{C_{22}C_{33}-C_{23}^2}\sqrt{C_{55}}} \\
\alpha_2&=\frac{C_{23}-C_{33}}{2C_{22}-3C_{23}+C_{33}} \\
\alpha_3&=\frac{C_{55}(2C_{22}-3C_{23}+C_{33})}{C_{44}C_{55}-C_{45}^2} \\
q_{uy}^{(2)}&=\frac{q_0x^4}{24 (C_{22}-2C_{23}+C_{33})}
\end{align}
and the constants $b_i$ ($i=1,2,\ldots,6$) are determined by boundary conditions (69)--(71).
\subsection{Micropolar ESL-FSDT beam model}
The general solution to Eqs.~(45)--(47) for uniformly distributed load $q=q_0$ and moment $m=m_0$ is
\begin{align}
u_y&=d_1-d_2x-\frac{d_3x^2}{2}-d_4\left(\frac{x^3}{3}-\beta_3x\right)-\frac{\beta_2}{\beta_1}\left(d_5e^{\beta_1x}-d_6e^{-\beta_1x}\right)\nonumber \\ &+q_{uy}^{(2)}-\frac{\beta_3}{2}\frac{\partial^2 q_{uy}^{(2)}}{\partial x^2}+m_{\psi z}^{(1)} \\
\psi_z&=-d_2-d_3x-d_4\left(x^2+\beta_4\right)-\frac{C_{22}-C_{23}}{C_{23}-C_{33}} \left(d_5e^{\beta_1x}+d_6e^{-\beta_1x}\right)\nonumber \\ &+\frac{\partial q_{uy}^{(2)}}{\partial x}+\frac{\beta_4}{2}\frac{\partial^3 q_{uy}^{(2)}}{\partial x^3}+\frac{C_{44}-C_{55}}{2(C_{45}-C_{55})}\frac{\partial m_{\psi z}^{(1)}}{\partial x} \\
\phi_x&=d_2+d_3x+d_4x^2+d_5e^{\beta_1 x}+d_6e^{-\beta_1 x}-\frac{\partial q_{uy}^{(2)}}{\partial x} 
\end{align}
where
\begin{align}
\beta_1&=\frac{\alpha_1\sqrt{C_{55}}}{\sqrt{C_{44}+2C_{45}+C_{55}}} \\
\beta_2&=\frac{C_{23}(C_{44}-2C_{45}-3C_{55})+C_{33}(C_{55}-C_{44})+2C_{22} (C_{45}+C_{55})}{(C_{23}-C_{33})(C_{44}+2C_{45}+C_{55})} \\
\beta_3&=\frac{C_{23}(C_{45}-3C_{55})+2C_{22}C_{55}+C_{33}(C_{55}-C_{45})}{C_{44}C_{55}-C_{45}^2} \\
\beta_4&=\beta_2\frac{(C_{23}-C_{33})(C_{44}+2C_{45}+C_{55})}{2\left(C_{45}^2-C_{44}C_{55}\right)} \\
 m_{\psi z}^{(1)}&=\frac{(C_{45}-C_{55})m_0x}{2(C_{44}C_{55}-C_{45}^2)}
\end{align}
The general solutions (with $q_0=m_0=0$) can be used to formulate the shape functions for nodally-exact beam finite elements; for details in the micropolar case, see \citep{karttunen2019a}.
\subsection{Microrotation and macrorotation relationships}
In Section 2.3, the second-order macrorotation-gradient was approximated by a central difference formula within a unit cell to establish relationships between the microrotation and the macrorotation and their gradients. Moreover, the third-order macrorotation-gradient that would have considerably affected the boundary conditions of the resulting beam theory was removed from micropolar beam model derivation that followed. Now, without considering any boundary conditions and using the micropolar beam solution (94)--(96), we obtain on the basis of Eq.~(16)
\begin{equation}
\frac{\partial \psi_z}{\partial x}-\frac{\partial \omega_z}{\partial x}=A_1q_0+A_2\left(d_5e^{\beta_1x}-d_6e^{-\beta_1x}\right) 
\end{equation}
where $A_1$ and $A_2$ are a constants. Equation (102) implies that the unit cell approximation (16) holds for the polynomial part of the homogeneous beam solution but not for the boundary layer described by the exponential terms. It follows that Eq.~(16) should provide a reasonable approximation especially for slender beams with distance from the boundary layer in the absence of distributed loads. Furthermore, the difference between the microrotation and macrorotation is
\begin{equation}
\psi_z-\omega_z=A_3m_0+A_4q_0x+A_5d_4+A_6\left(d_5e^{\beta_1x}+d_6e^{-\beta_1x}\right)    
\end{equation}
where $A_j$ $(j=3,4,5,6)$ are constants. Result (103) suggests that in addition to the boundary layer, there may be a constant difference between the rotations even for $q=m=0$. 
\section{Numerical examples and discussion}
We study the static bending of the micropolar, couple-stress and classical beam models presented in Section 3. First we consider a stretching-dominated X-core beam with and without the sandwich face sheets. For the remainder of this section, the focus will be on bending-dominated lattice core sandwich beams. In relation to the previous Section 4.5, we will show that the boundary effects vanish from a simply-supported micropolar beam under a uniformly distributed moment $m_0$. Moreover, this case will be used as the basis for illustrating pure symmetric and antisymmetric shear deformations of a web-core lattice beam. The solutions to the 1-D beam problems are obtained using the general solutions derived in Section 4, whilst 2-D reference solutions are computed using Euler--Bernoulli FE beam frames modeled by cubic B23 elements in Abaqus. These elements are nodally-exact and only one element is required for each span in the present bending problems. Note that the constituents of a 2-D FE beam frame do not exhibit any significant shear strains, but when the 2-D problem is reduced to a 1-D ESL beam problem, the shear behavior needs to be considered.
\subsection{X-core beam under a mid-point load}
We model a clamped-clamped X-core beam under a mid-point load $2F$ by its symmetric half. The unit cell length, height and width are $l=h=b=0.1$ m, respectively, and the lattice member thicknesses are $t_d=t_f=3$ mm for the diagonals and face sheets, respectively. The webs are excluded from the model. The members are made of steel $(E=210$ GPa, $\nu=0.3)$. The boundary conditions, e.g., for the micropolar beam are
\begin{equation}
\begin{aligned}
x&=0:\ \phi_x=\psi_z=u_y=0 \\
x&=L:\ \phi_x=\psi_z=0,\ Q_{xy}=F
\end{aligned}
\end{equation}
which are used to solve constants $d_i$ ($i=1,2,\ldots,6$) in Eqs.~(94)--(96).

In Fig.~5, the 1-D beam models are compared to the corresponding 2-D FE beam frame in terms of the maximum transverse deflections for different beam lengths. In Fig.~5a, the X-core beam has no face sheets and only the 1-D couple-stress and micropolar models converge toward to the 2-D solution. The classical beam models give erroneous results, however, it can be seen from the classical cases that the 1-D beams do not exhibit any significant shear deformations as the classical EBT and FSDT (Timoshenko) models give the same results. In Fig.~5b, the faces have been added to the beam and the shear deformations play a notable role since the classical EBT model is rather stiff compared to the shear deformable models. The differences between the FSDT beam models are small. 

\begin{figure}
\centering
\includegraphics[scale=0.42]{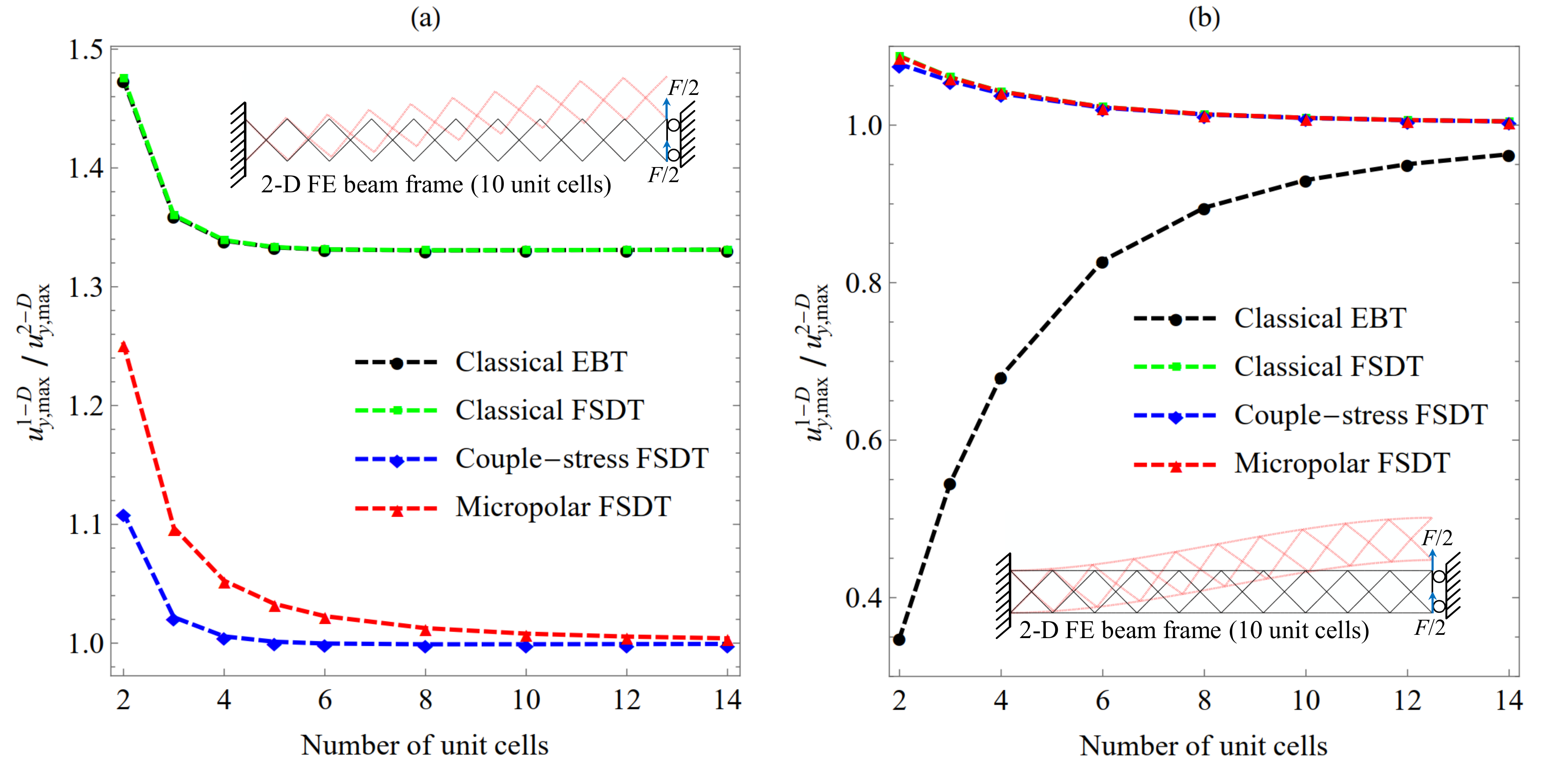}
\caption{a) Bending of an X-core beam without sandwich face sheets. The classical 1-D models are too flexible. b) Bending of an X-core beam with face sheets. The FSDT models give practically identical results. It follows that the classical FSDT model is sufficent for modeling typical stretching-dominated sandwich beams with lattice cores.}
\end{figure}

The difference between the classical and non-classical results in Fig.~5a can be explained by looking at the local/global bending stiffness ratio $C_{33}/C_{22}$. For $h=l$, Eq.~(56) gives for the couple-stress and micropolar beams
\begin{equation}
\frac{C_{33}}{C_{22}}=\frac{2l^2+t_d^2}{6l^2}\quad\rightarrow\quad \frac{C_{33}}{C_{22}}=\frac{1}{3}\quad \textrm{for}\quad t_d<<l    
\end{equation}
which shows that the contribution of the local bending stiffness to the total bending stiffness of the beam is considerable in the absence of the face sheets; for $t_d<<l$, $C_{33}=4\sqrt{2}EI_d$. In fact, it can be seen in Fig.~5a that the classical models are in error by 33\% as essentially indicated by Eq.~(105) because they consider only the global bending behavior. In more detail, the classical 1-D models are too flexible and overestimate the deflections due to the fact that they were obtained in Section 3 by assuming that they have no internal local moment ($P_{xz}=0$) to capture the local bending behavior which would stiffen the beam. Nevertheless, this is not a bad assumption for typical stretching-dominated sandwich beams where the face sheets are present. For example, in Fig.~5b we have $C_{33}/C_{22}=1/874$ for the local/global stiffness ratio and the global bending dominates over the local one. Increasing the face thickness increases $C_{33}$ but the $C_{33}/C_{22}$-ratio remains small for physically reasonable face thicknesses.

\begin{figure}
\centering
\includegraphics[scale=0.42]{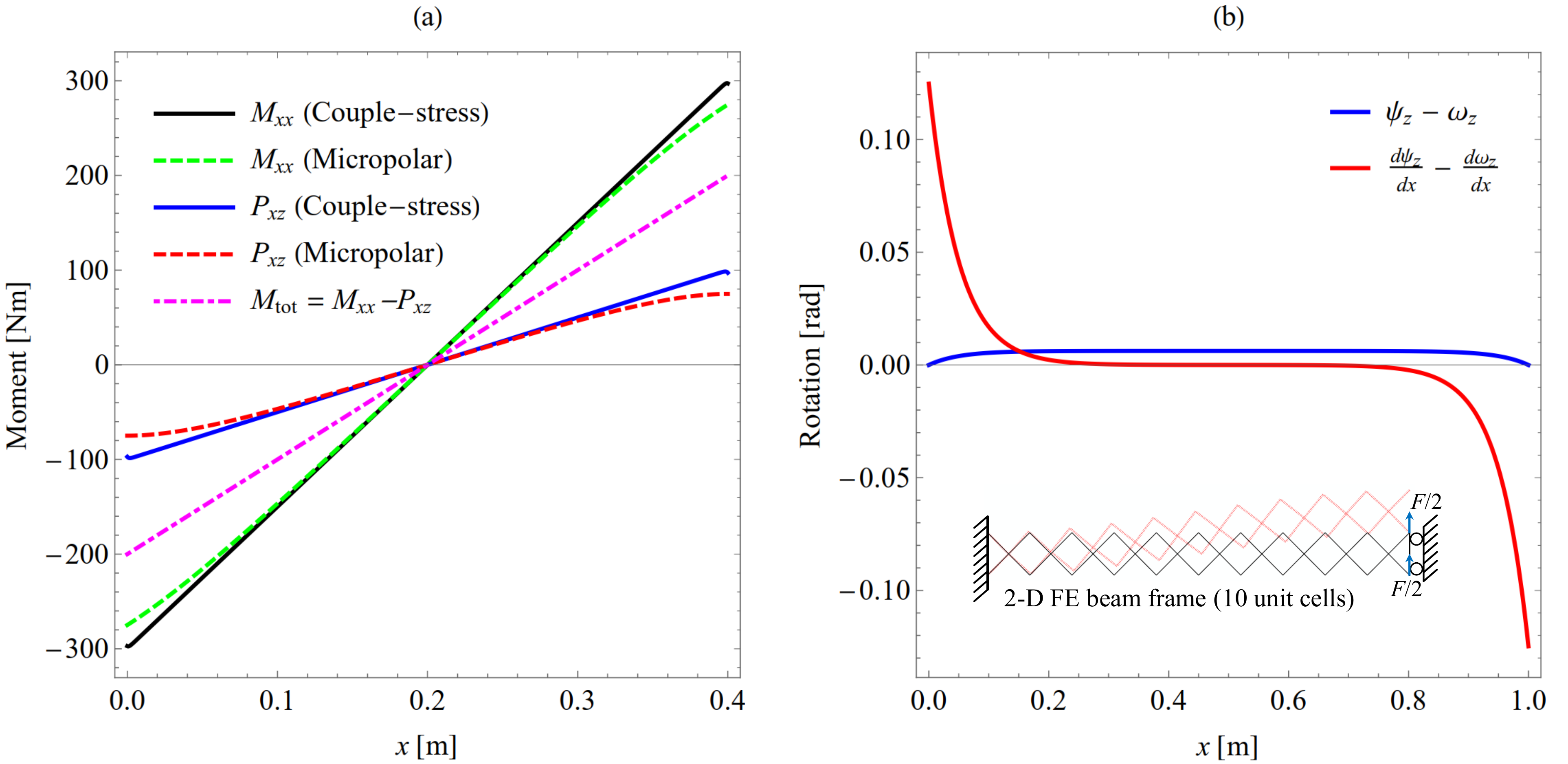}
\caption{a) Bending moment diagrams for X-core couple-stress and micropolar beams (no face sheets). b) Differences between the microrotation $\psi_z$ and macrorotation $\omega_z$ and their first derivatives in the micropolar model.}
\end{figure}

To take a closer look at the differences between the two non-classical models, Fig.~6a presents the couple-stress and micropolar bending moments for a faceless X-core beam which is four unit cells long. Both models give the same total bending moment but there are differences in the global $M_{xx}$ and local $P_{xz}$ parts. In the couple-stress model the moments are linear along the beam except for small notches at the ends, whereas in the micropolar case the boundary effects are more prominent. The homogenized 1-D couple-stress model appears to predict the periodic 2-D behavior more accurately for shorter faceless beams (Fig.~5a). It is noteworthy that since there is no shear deformation in this problem, the split of shear into symmetric and antisymmetric parts in the micropolar beam does not play a role here.

Figure~6b considers the micropolar beam model and the differences between the microrotation and macrorotation and their first derivatives. Since $q_0=0$, there is no difference between the first derivatives with distance from the boundaries, as predicted by Eq.~(102). The rotations $\psi_z$ and $\omega_z$ are zero at the edges but there is a constant difference between them with distance from the boundaries as suggested by the general solution through Eq.~(103).

When it comes to stretching-dominated lattice core beams it can be concluded that:
\begin{itemize}
    \item Without the face sheets, a lattice core beam does not exhibit the sandwich effect, i.e., bending is not resisted by the face sheet membrane action. In such a case, the local bending related to the internal rotations of the lattice beam can contribute considerably to the beam deflections and non-classical beam models are needed for accurate results.
    \item With the face sheets, a classical FSDT model is typically enough for modeling stretching-dominated sandwich beams with lattice cores where the core members carry the shear as mainly as axial rods. As the shear deformations are not neglible, the classical Euler-Bernoulli ESL model can be too stiff for such sandwich beams unless they are very slender.
    \item Although not studied here in detail, local loads or nonlinear deformations can make ideally stretching-dominated core members behave as beams making the core bending-dominated especially in the presence of imperfections due to manufacturing (e.g. initial curvature). 
\end{itemize}
\subsection{Simply-supported micropolar beam under a uniformly distributed moment}
We consider a simply-supported web-core beam without the X-braces subjected to a uniformly distributed moment $m_0=10000$ Nm/m. The unit cell length, height and width are $l=h=b=0.1$ m, respectively, and the thicknesses are $t_f=t_w=3$ mm for the faces and webs, respectively. The members are made of steel $(E=210$ GPa, $\nu=0.3)$ and the face-web joints are rigid [$\Theta=3EI_wl/h$, Eq.~(55)]. The boundary conditions for the beam ends are $u_y=M_{xx}=P_{xz}=0$ and the analytical solution to the problem is
\begin{equation}
u_y=0, \quad \psi_z=-\phi_x=m_0l^2/(24EI_f)   
\end{equation}
It can be seen that the beam does not exhibit any boundary effects. We consider a beam which is one unit cell long and substitute the 1-D micropolar results into the corresponding 2-D FE beam frame (where the web stiffnesses are not halved as in Fig.~4) to obtain the localized 1-D response.

\begin{figure}
\centering
\includegraphics[scale=0.5]{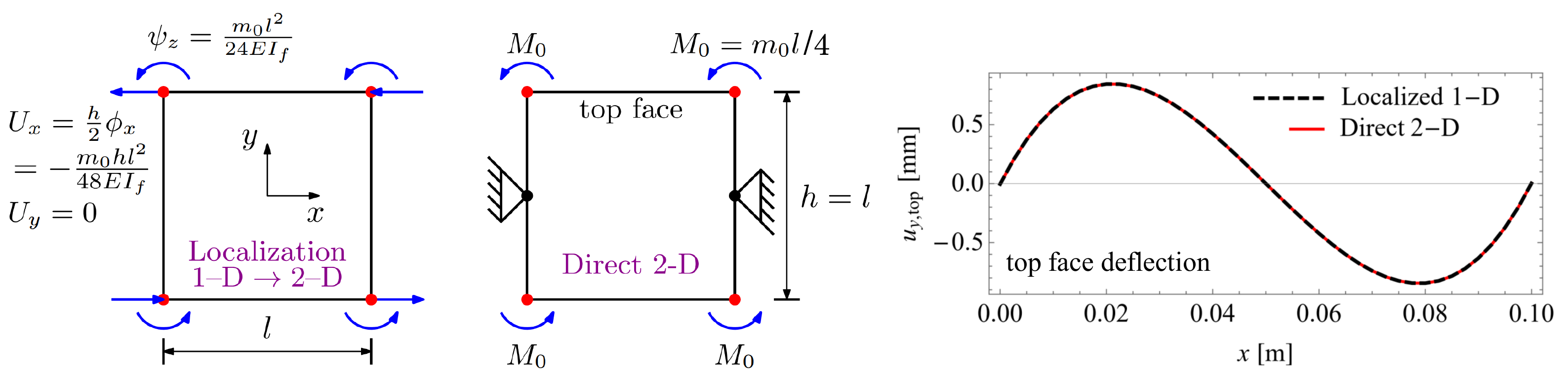}
\caption{1-D displacements and rotations are substituted into a 2-D beam frame that corresponds to the 1-D beam (left). Corresponding direct 2-D analysis is carried out by nodal loads and boundary conditions equivalent to the 1-D case. Both approaches give the same final displacement field as illustrated by the top face transverse deflection. In other words, the micropolar 1-D homogenization-localization procedure that employs approximate FSDT kinematics leads to the same result as the full 2-D FE system. This implies that the microrotation provided by the micropolar theory is a good way to describe the internal rotations of a lattice core.}
\end{figure}

Figure 7 shows that in the absence of boundary effects, the localized 1-D model is equivalent to the corresponding 2-D model even when the macrostructure (beam) and microstructure (unit cell) are of the same length ($L=l$). Therefore, the requirement of scale separation ($L>>l$) typical to second-order computational homogenization methods, which consider the local bending effects as well, appears to become irrelevant in this special case  \citep{kouznetsova2002,larsson2007,geers2010,matouvs2017}. In the presence of boundary effects, however, it has been shown that there are some differences between localized 1-D micropolar beam and 2-D FE beam frame solutions \citep{karttunen2019a}.

Figure 8a presents the result of the 1-D-to-2-D localization of Fig.~7 in full. In this case of horizontal shear deformation, the symmetric (25) and antisymmetric (26) shear strains are equal ($\gamma_s=\gamma_a$). Figs.~8b--d present other examples of how the shear deformations manifest when 1-D results (constant rotations) are substituted into the corresponding 2-D FE beam frame. Fig.~8b shows a case of vertical shear. This type of a deformation will become dominant, for example, in Fig.~1 if the webs are a lot thicker than the face sheets ($t_w>>t_f$). Pure symmetric shear appears in Fig.~8c when the cross sectional rotation coincides with the slope and the microrotation is zero. Figure~8d displays the non-classical pure antisymmetric shear deformation which is accounted for only by the micropolar beam model. Finally, we note that the localization of 1-D micropolar beam results to calculate the periodic stresses of web-core sandwich beams has been carried out and discussed earlier at length in \citep{karttunen2019a}.

\begin{figure}
\centering
\includegraphics[scale=0.48]{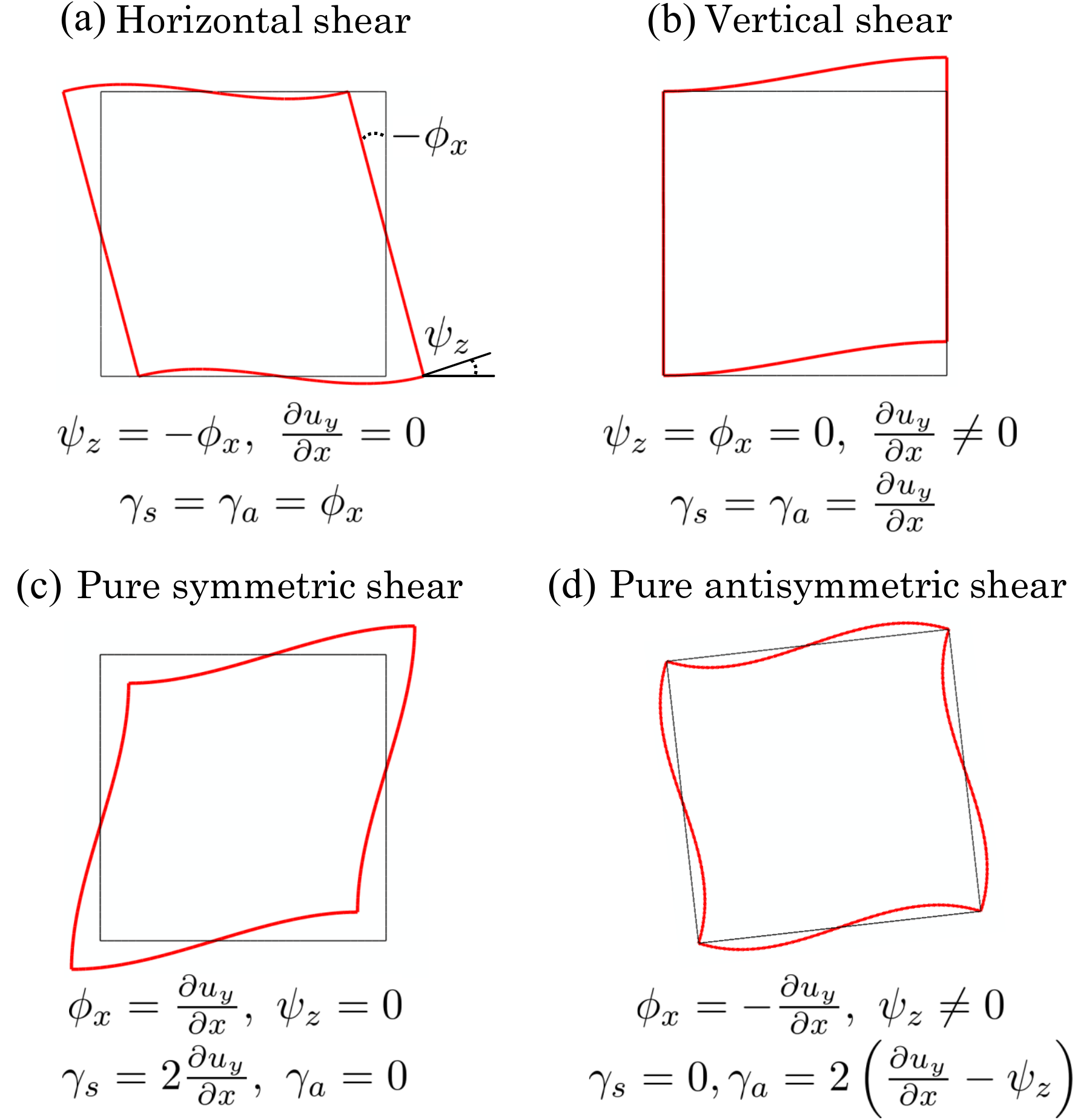}
\caption{Different types of shear deformations and how they manifest when 1-D micropolar constant rotations are substituted into a corresponding 2-D FE web-core beam frame. Subfigure (a) corresponds to the case in Figure 7.}
\end{figure}
\subsection{Cantilever web-core beam}
An all-steel $(E=210$ GPa, $\nu=0.3)$ cantilever web-core beam with a tip point load $F=100$ N is considered here. The unit cell width, length and height are $b=0.05$ m, $l=0.12$ m and $h=0.043$ m, respectively. The web thickness is $t_w=4$ mm and the face thickness is varied in the calculations. The face-web joints are rigid. Similar beams have been studied experimentally in three-point bending by \cite{karttunen2017b}. The boundary conditions for the free end of a micropolar beam are $M_{xx}=P_{xz}=0$ and $Q_{xy}=F$; for the clamped end, see Eq.~(104).

In Fig.~9a, the 1-D FSDT beam models are compared to the corresponding 2-D FE beam frame in terms of the maximum transverse deflections for three different face thicknesses ($t_f=1,3,6$ mm). Figure~9b shows the 2-D FE frames for the case $L=6l$. When the faces are thin ($t_f=1$ mm), the 1-D classical and micropolar FSDT beam models are in excellent agreement with the 2-D beam frame, whereas the couple-stress model is overtly stiff. Fig.~9b also presents the central axis deflection of the 2-D frame in an average sense. It can be seen that the slope at the clamped end is not zero for $t_f=1$ mm, however, the slope is zero in the couple-stress model due to the boundary condition $\omega_z=0$, see Eqs.~(71) and (5). This slope boundary condition has its origins in the assumption that the corner rotations of the unit cell of Fig.~2 are constrained to follow the macrorotation behavior according to Eq.~(10). The couple-stress beam is too stiff also for the face thickness $t_f=3$ mm, for which the 2-D frame still exhibits a similar, wavy deformation as for $t_f=1$ mm. For thicker face sheets ($t_f=6$ mm), the waviness diminishes, and at the clamped end the average central axis slope approaches zero in 2-D so that also the 1-D couple-stress model works for this case and is actually slightly more flexible for $t_f=6$ mm than the micropolar model. The couple-stress model was derived by assuming $Q_a=0$ in the micropolar case so that the shear deformations, as a whole, are allowed to occur more freely.
\begin{figure}
\centering
\includegraphics[scale=0.5]{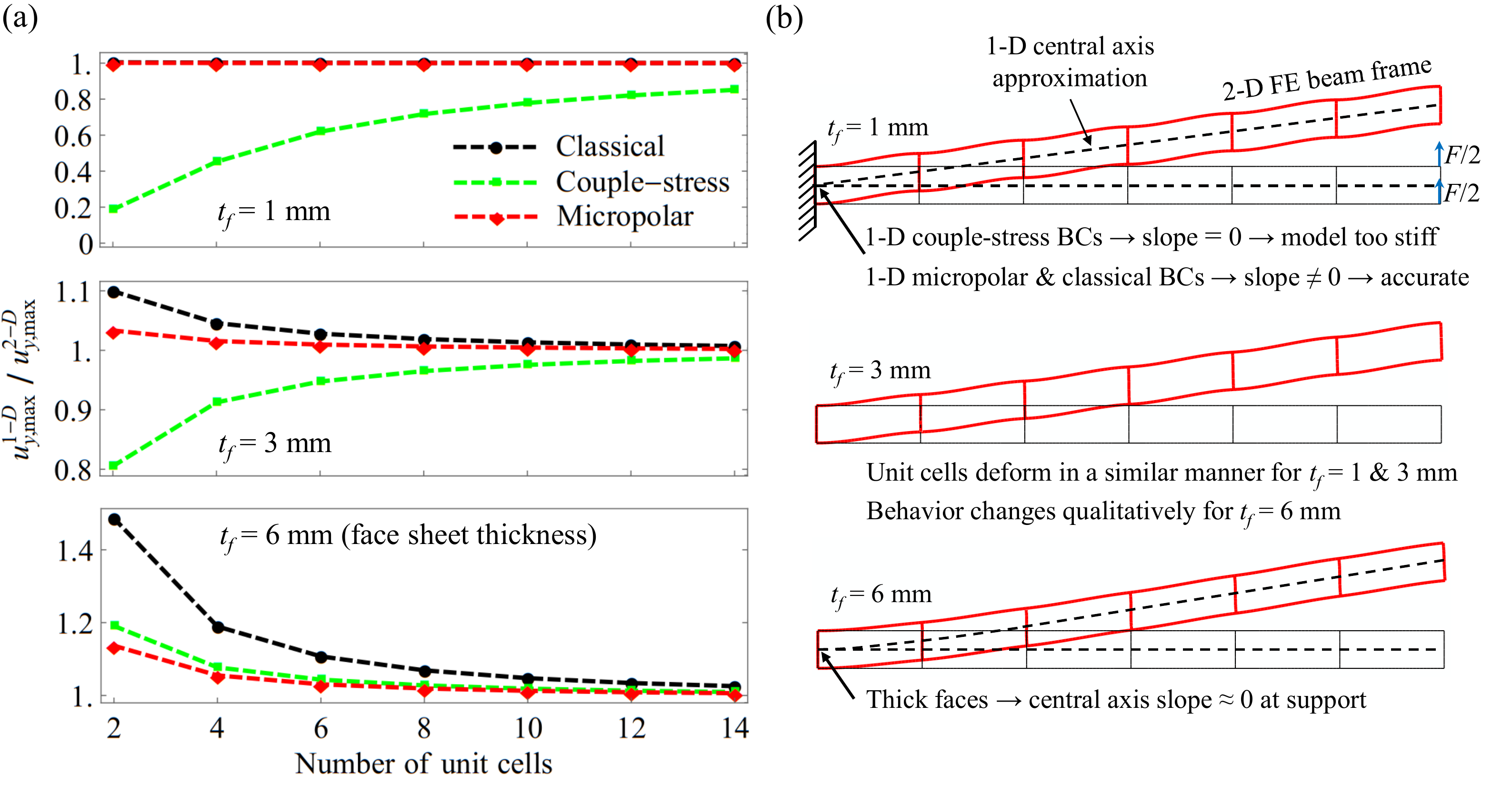}
\caption{Cantilever web-core beam under a tip point load. a) Comparison of 1-D and 2-D results. The couple-stress model is too stiff for thin faces as it requires the slope to be zero at a clamped end. b) 2-D FE deformations for a beam that is six unit cells long. The slope at the clamped end approaches zero as the face thickness increases.}
\end{figure}
\subsection{Web-core beam under uniformly distributed load}
As the final example, a clamped-simply-supported web-core beam subjected to a uniform line load $q=1000$ N/m is studied. The parameters are the same as in the previous case except that the face thickness is now fixed to $t_f=3$ mm and both rigid and flexible, laser-welded face-web joints with rotational stiffness $k_\theta=2675$ Nm are considered \citep{romanoff2007c}.

Figure~10a presents the deflections along the beam for the 1-D FSDT beam models and the 2-D frame when the face-web joints are rigid. It can be seen that the couple-stress beam is too stiff again at the clamped boundary and, in fact, the model predicts the location of the maximum transverse displacement incorrectly. Fig.~10b shows that as the joints become flexible, the face sheet waviness reduces and the prediction given by the couple-stress model improves. As the joint stiffness is reduced further, the slope at the clamped end approaches zero. 
\begin{figure}
\centering
\includegraphics[scale=0.51]{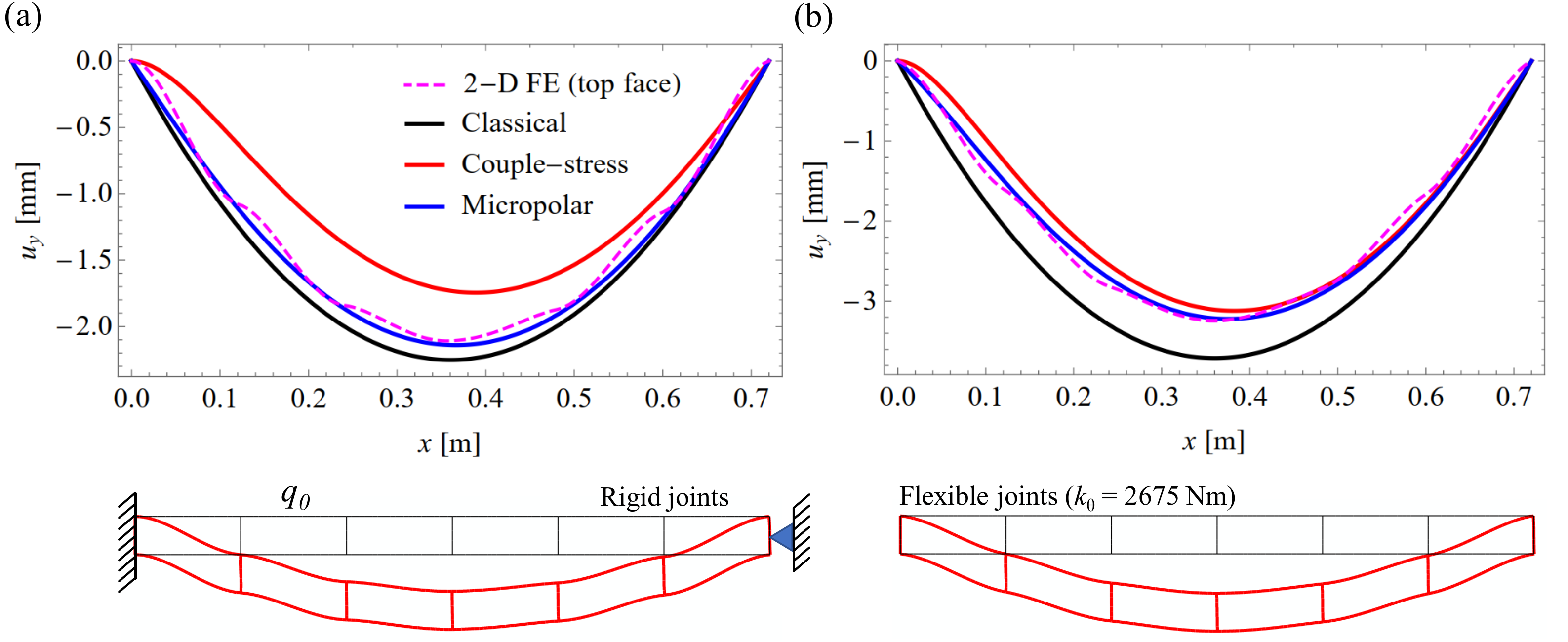}
\caption{Clamped-simply-supported web-core beam under a uniform line load $q_0=1000$ N/m. a) Rigid face-web joints. b) Flexible joints. As the joint stiffness decreases, the average central axis slope of the 2-D frame at the clamped edge approaches zero so that the prediction given by the couple-stress model improves.}
\end{figure}

In summary on bending-dominated lattice beams, we find that:
\begin{itemize}
    \item In the classical FSDT model, there is no rotation variable to describe the unit cell corner rotations (Fig.~2) and, consequently, there are no non-classical strain measures which are needed to capture the 2-D behavior accurately by a 1-D model; the lack of local bending and antisymmetric shear causes the 1-D classical model to be too flexible.
    \item In the couple-stress FSDT model, the unit cell corner rotations are constrained to follow the macrorotation $\omega_z$. This model works for bending-dominated lattice core sandwich beams when the slope boundary condition $\partial u_y/\partial x=0$ set in the couple-stress model through $\omega_z=0$ corresponds to behavior of the actual sandwich beam (a 2-D FE beam frame in this paper). The slope behavior is critical in the vicinity of any discontinuities, including boundaries as well as point loads and moments. In terms of strains, the couple-stress model cannot capture the antisymmetric shear which can be seen first schematically in Fig.~8d and then in Figs.~9b and 10 from the 2-D results.
    \item In the micropolar FSDT beam model, the microrotation provides an accurate description for the unit cell corner rotations and the 1-D beam slope adjusts to the average central axis slope of a 2-D FE beam frame. The micropolar model shows that both the local curvature and antisymmetric shear strain are essential features of a 1-D ESL beam model for lattice core sandwich panels.
\end{itemize}
\section{Conclusions}
In this paper, a 1-D micropolar equivalent single-layer (ESL) beam model based on the first-order shear deformation theory (FSDT) was derived for sandwich panels with lattice cores. The model was reduced to a couple-stress ESL-FSDT and two classical ESL beam models. The four models were used to investigate the bending of lattice core sandwich beams to point out differences between the models and to make it easier for practitioners to choose a proper beam model for a particular engineering application. In short, 
\begin{itemize}
    \item The 4th-order classical ESL-FSDT beam is a good first choice for modeling \textit{stretching-dominated} beams.
    \item The 6th-order micropolar ESL-FSDT beam provides reliable results for both stretching- and \textit{bending-dominated} sandwich structures.
    \item The 6th-order couple-stress model is often too stiff and does not offer any significant benefits over the micropolar model when modeling sandwich beams that have face sheets. 
\end{itemize}
The accuracy of the obtained micropolar results also suggests that a third-order shear deformation model (TSDT) may not provide any significant benefits over the developed FSDT model. 

The classical and micropolar ESL-FSDT approaches have been extended to 2-D sandwich plates with lattice cores  \citep{karttunen2019b,nampally2020}. In the latter plate paper, it was shown that 2-D micropolar plate finite elements provide computationally cost-effective solutions to lattice core sandwich panel problems. For example, in a nonlinear bending problem, the number of 2-D micropolar plate elements required to model the bending of a lattice core panel was of the order $10^3$, whereas for the corresponding 3-D reference model the order was $10^5$.

The micropolar beam formulation was based on the fact that the second-order macrorotation-gradient and the micropolar antisymmetric shear deformation describe the same kind of strain; for a graphical presentation of the former see \citep{shaat2016} and for the latter see Fig.~8d above and \citep{debellis2011}. On a related note, by studying how a particular discrete lattice deforms under different loads and boundary conditions, one can identify the strain measures that are needed to model the discrete lattice as a continuum material. Digital image correlation, for example, provides methods for the experimental identification of higher-order displacement gradients and tools to visualize them \citep{lu2000}.

The discrete unit cell, which is subjected to the identified strains whatever the chosen homogenization method, is very much the key to successful modeling of lattice materials. If an open cell such as the one depicted in Fig.~1 had been chosen for the web-core instead of the closed rectangular cell in Fig.~4, we would have $\Theta=0$ for the coupling term of Eq.~(55) between the local and global moments since the moments do not interact via the web in the open cell. That is to say, the unit cell choice has an effect on the constitutive matrix and it is good practice to consider a few different options even for the same lattice material. Nevertheless, in the present framework, the general form of the micropolar governing equations (44)--(51) is always the same for the four-node representation of Fig.~2a and only the constitutive matrix (35) changes between different unit cell choices and lattice materials. So far the micropolar two-scale constitutive modeling approach has been applied successfully to several types of cores, see \cite{nampally2019} for Y-core as well as honeycomb and corrugated cores. At the moment, the applicability of the two-scale method to, for example, a chiral or a general core remains an open question.

If we were to add more nodes to the general unit cell of Fig.~2a, displacement and rotation gradients of higher order than those considered in (8)--(10) would also contribute to the discrete-to-continuum transformation (27) without rendering the constitutive behavior unstable. Simply put, more nodes means more gradients, as was basically shown by \cite{noor1980}. Again, the need for the additional nodes and strain gradients depends on the types of deformations a lattice material exhibits in the intended applications. By increasing the number of nodes in the unit cell of Fig.~2a, it follows that the two-scale constitutive modeling method used in this paper is not limited to the discussed beam models but may be used with appropriate modifications to formulate, for example, strain gradient lattice beam and plate models similar to those developed by \cite{khakalo2019,khakalo2020}.
\section*{Acknowledgements}
The first author acknowledges the funding received from the European Union's Horizon 2020 research and innovation programme under the Marie Sk\l{}odowska--Curie grant agreement No 745770. The financial support is greatly appreciated.
The second author gratefully acknowledges the support of his research through the Oscar S. Wyatt Endowed Chair. The authors also acknowledge CSC -- IT Center for Science, Finland, for computational resources (Abaqus usage).
\section*{Declaration of competing interests}
The authors declare that they have no known competing financial interests or personal relationships that could have appeared to influence the work reported in this paper. 
\bibliographystyle{elsarticle-harv}
\bibliography{microlitera}

\begin{thebibliography}{80}
\expandafter\ifx\csname natexlab\endcsname\relax\def\natexlab#1{#1}\fi
\providecommand{\url}[1]{\texttt{#1}}
\providecommand{\href}[2]{#2}
\providecommand{\path}[1]{#1}
\providecommand{\DOIprefix}{doi:}
\providecommand{\ArXivprefix}{arXiv:}
\providecommand{\URLprefix}{URL: }
\providecommand{\Pubmedprefix}{pmid:}
\providecommand{\doi}[1]{\href{http://dx.doi.org/#1}{\path{#1}}}
\providecommand{\Pubmed}[1]{\href{pmid:#1}{\path{#1}}}
\providecommand{\bibinfo}[2]{#2}
\ifx\xfnm\relax \def\xfnm[#1]{\unskip,\space#1}\fi
\bibitem[{Allen(1969)}]{allen1969}
\bibinfo{author}{Allen, H.G.}, \bibinfo{year}{1969}.
\newblock \bibinfo{title}{Analysis and Design of Structural Sandwich Panels}.
\newblock \bibinfo{publisher}{Pergamon Press}.
\bibitem[{andrew and Jasiuk(2006)}]{yoo2006}
\bibinfo{author}{andrew, Y.}, \bibinfo{author}{Jasiuk, I.},
  \bibinfo{year}{2006}.
\newblock \bibinfo{title}{Couple-stress moduli of a trabecular bone idealized
  as a 3{D} periodic cellular network}.
\newblock \bibinfo{journal}{J. Biomech.} \bibinfo{volume}{39},
  \bibinfo{pages}{2241--2252}.
\bibitem[{Asghari et~al.(2011)Asghari, Rahaeifard, Kahrobaiyan and
  Ahmadian}]{asghari2011}
\bibinfo{author}{Asghari, M.}, \bibinfo{author}{Rahaeifard, M.},
  \bibinfo{author}{Kahrobaiyan, M.H.}, \bibinfo{author}{Ahmadian, M.T.},
  \bibinfo{year}{2011}.
\newblock \bibinfo{title}{The modified couple stress functionally graded
  {T}imoshenko beam formulation}.
\newblock \bibinfo{journal}{Mater. Design} \bibinfo{volume}{32},
  \bibinfo{pages}{1435--1443}.
\bibitem[{Ashby(2006)}]{ashby2006}
\bibinfo{author}{Ashby, M.}, \bibinfo{year}{2006}.
\newblock \bibinfo{title}{The properties of foams and lattices}.
\newblock \bibinfo{journal}{Philos. T. R. Soc. A.} \bibinfo{volume}{364},
  \bibinfo{pages}{15--30}.
\bibitem[{Bacigalupo and Gambarotta(2014)}]{bacigalupo2014}
\bibinfo{author}{Bacigalupo, A.}, \bibinfo{author}{Gambarotta, L.},
  \bibinfo{year}{2014}.
\newblock \bibinfo{title}{Homogenization of periodic hexa-and tetrachiral
  cellular solids}.
\newblock \bibinfo{journal}{Compos. Struct.} \bibinfo{volume}{116},
  \bibinfo{pages}{461--476}.
\bibitem[{Bacigalupo and Gambarotta(2019)}]{bacigalupo2019}
\bibinfo{author}{Bacigalupo, A.}, \bibinfo{author}{Gambarotta, L.},
  \bibinfo{year}{2019}.
\newblock \bibinfo{title}{Generalized micropolar continualization of 1{D} beam
  lattices}.
\newblock \bibinfo{journal}{Int. J. Mech. Sci.} \bibinfo{volume}{155},
  \bibinfo{pages}{554--570}.
\bibitem[{Ba{\v{z}}ant and Christensen(1972)}]{bazant1972}
\bibinfo{author}{Ba{\v{z}}ant, Z.P.}, \bibinfo{author}{Christensen, M.},
  \bibinfo{year}{1972}.
\newblock \bibinfo{title}{Analogy between micropolar continuum and grid
  frameworks under initial stress}.
\newblock \bibinfo{journal}{Int. J. Solids. Struct.} \bibinfo{volume}{8},
  \bibinfo{pages}{327--346}.
\bibitem[{Biagi and Bart-Smith(2007)}]{biagi2007}
\bibinfo{author}{Biagi, R.}, \bibinfo{author}{Bart-Smith, H.},
  \bibinfo{year}{2007}.
\newblock \bibinfo{title}{Imperfection sensitivity of pyramidal core sandwich
  structures}.
\newblock \bibinfo{journal}{Int. J. Solids Struct.} \bibinfo{volume}{44},
  \bibinfo{pages}{4690--4706}.
\bibitem[{Birman and Kardomateas(2018)}]{birman2018}
\bibinfo{author}{Birman, V.}, \bibinfo{author}{Kardomateas, G.A.},
  \bibinfo{year}{2018}.
\newblock \bibinfo{title}{Review of current trends in research and applications
  of sandwich structures}.
\newblock \bibinfo{journal}{Compos. Part B-Eng.,} .
\bibitem[{Bright and Smith(2007)}]{bright2007}
\bibinfo{author}{Bright, S.R.}, \bibinfo{author}{Smith, J.W.},
  \bibinfo{year}{2007}.
\newblock \bibinfo{title}{A new design for steel bridge decks using laser
  fabrication}.
\newblock \bibinfo{journal}{Struct. Eng.,} \bibinfo{volume}{85}.
\bibitem[{Chen and Huang(2019)}]{chen2019}
\bibinfo{author}{Chen, W.}, \bibinfo{author}{Huang, X.}, \bibinfo{year}{2019}.
\newblock \bibinfo{title}{Topological design of 3{D} chiral metamaterials based
  on couple-stress homogenization}.
\newblock \bibinfo{journal}{J. Mech. Phys. Solids} \bibinfo{volume}{131},
  \bibinfo{pages}{372--386}.
\bibitem[{Chen and Lui(2005)}]{chen2005}
\bibinfo{author}{Chen, W.H.}, \bibinfo{author}{Lui, E.M.},
  \bibinfo{year}{2005}.
\newblock \bibinfo{title}{Handbook of Structural Engineering}.
\newblock \bibinfo{publisher}{CRC Press}.
\bibitem[{Chen et~al.(2014)Chen, Liu and Hu}]{chen2014}
\bibinfo{author}{Chen, Y.}, \bibinfo{author}{Liu, X.}, \bibinfo{author}{Hu,
  G.}, \bibinfo{year}{2014}.
\newblock \bibinfo{title}{Micropolar modeling of planar orthotropic rectangular
  chiral lattices}.
\newblock \bibinfo{journal}{Cr. Mecanique} \bibinfo{volume}{342},
  \bibinfo{pages}{273--283}.
\bibitem[{Chowdhury and Reddy(2019)}]{chowdhury2019}
\bibinfo{author}{Chowdhury, S.R.}, \bibinfo{author}{Reddy, J.N.},
  \bibinfo{year}{2019}.
\newblock \bibinfo{title}{Geometrically exact micropolar {T}imoshenko beam and
  its application in modelling sandwich beams made of architected lattice
  core}.
\newblock \bibinfo{journal}{Compos. Struct.} \bibinfo{volume}{226, 111228}.
\bibitem[{Chung and Waas(2009)}]{chung2009}
\bibinfo{author}{Chung, J.}, \bibinfo{author}{Waas, A.M.},
  \bibinfo{year}{2009}.
\newblock \bibinfo{title}{The micropolar elasticity constants of circular cell
  honeycombs}.
\newblock \bibinfo{journal}{Philos. T. R. Soc. A.} \bibinfo{volume}{465},
  \bibinfo{pages}{25--39}.
\bibitem[{De~Bellis and Addessi(2011)}]{debellis2011}
\bibinfo{author}{De~Bellis, M.L.}, \bibinfo{author}{Addessi, D.},
  \bibinfo{year}{2011}.
\newblock \bibinfo{title}{A {C}osserat based multi-scale model for masonry
  structures}.
\newblock \bibinfo{journal}{Int. J. Multiscale Com.} \bibinfo{volume}{9},
  \bibinfo{pages}{543--563}.
\bibitem[{Dos~Reis and Ganghoffer(2012)}]{dos2012}
\bibinfo{author}{Dos~Reis, F.}, \bibinfo{author}{Ganghoffer, J.F.},
  \bibinfo{year}{2012}.
\newblock \bibinfo{title}{Construction of micropolar continua from the
  asymptotic homogenization of beam lattices}.
\newblock \bibinfo{journal}{Comput. Struct.} \bibinfo{volume}{112},
  \bibinfo{pages}{354--363}.
\bibitem[{Duan et~al.(2018)Duan, Wen and Fang}]{duan2018}
\bibinfo{author}{Duan, S.}, \bibinfo{author}{Wen, W.}, \bibinfo{author}{Fang,
  D.}, \bibinfo{year}{2018}.
\newblock \bibinfo{title}{A predictive micropolar continuum model for a novel
  three-dimensional chiral lattice with size effect and tension-twist coupling
  behavior}.
\newblock \bibinfo{journal}{J. Mech. Phys. Solids} \bibinfo{volume}{121},
  \bibinfo{pages}{23--46}.
\bibitem[{Eringen(1966)}]{eringen1966}
\bibinfo{author}{Eringen, A.C.}, \bibinfo{year}{1966}.
\newblock \bibinfo{title}{Linear theory of micropolar elasticity}.
\newblock \bibinfo{journal}{J. Math. Mech.} , \bibinfo{pages}{909--923}.
\bibitem[{Eringen(2012)}]{eringen2012}
\bibinfo{author}{Eringen, A.C.}, \bibinfo{year}{2012}.
\newblock \bibinfo{title}{Microcontinuum Field Theories: I. Foundations and
  Solids}.
\newblock \bibinfo{publisher}{Springer Science \& Business Media}.
\bibitem[{Ganghoffer et~al.(2018)Ganghoffer, Goda, Novotny, Rahouadj and
  Sokolowski}]{ganghoffer2018}
\bibinfo{author}{Ganghoffer, J.F.}, \bibinfo{author}{Goda, I.},
  \bibinfo{author}{Novotny, A.A.}, \bibinfo{author}{Rahouadj, R.},
  \bibinfo{author}{Sokolowski, J.}, \bibinfo{year}{2018}.
\newblock \bibinfo{title}{Homogenized couple stress model of optimal auxetic
  microstructures computed by topology optimization}.
\newblock \bibinfo{journal}{ZAMM-Z. Angew. Math. Me.} \bibinfo{volume}{98},
  \bibinfo{pages}{696--717}.
\bibitem[{Geers et~al.(2010)Geers, Kouznetsova and Brekelmans}]{geers2010}
\bibinfo{author}{Geers, M.G.D.}, \bibinfo{author}{Kouznetsova, V.G.},
  \bibinfo{author}{Brekelmans, W.A.M.}, \bibinfo{year}{2010}.
\newblock \bibinfo{title}{Multi-scale computational homogenization: Trends and
  challenges}.
\newblock \bibinfo{journal}{J. Comput. Appl. Math.} \bibinfo{volume}{234},
  \bibinfo{pages}{2175--2182}.
\bibitem[{Gesualdo et~al.(2017)Gesualdo, Iannuzzo, Penta and
  Pucillo}]{gesualdo2017}
\bibinfo{author}{Gesualdo, A.}, \bibinfo{author}{Iannuzzo, A.},
  \bibinfo{author}{Penta, F.}, \bibinfo{author}{Pucillo, G.P.},
  \bibinfo{year}{2017}.
\newblock \bibinfo{title}{Homogenization of a {V}ierendeel girder with elastic
  joints into an equivalent polar beam}.
\newblock \bibinfo{journal}{J. Mech. Mater. Struct.} \bibinfo{volume}{12},
  \bibinfo{pages}{485--504}.
\bibitem[{Goda and Ganghoffer(2015)}]{goda2015}
\bibinfo{author}{Goda, I.}, \bibinfo{author}{Ganghoffer, J.F.},
  \bibinfo{year}{2015}.
\newblock \bibinfo{title}{Identification of couple-stress moduli of vertebral
  trabecular bone based on the 3{D} internal architectures}.
\newblock \bibinfo{journal}{J. Mech. Behav. Biomed.} \bibinfo{volume}{51},
  \bibinfo{pages}{99--118}.
\bibitem[{Goncalves et~al.(2019)Goncalves, Karttunen and
  Romanoff}]{goncalves2019}
\bibinfo{author}{Goncalves, B.R.}, \bibinfo{author}{Karttunen, A.T.},
  \bibinfo{author}{Romanoff, J.}, \bibinfo{year}{2019}.
\newblock \bibinfo{title}{A nonlinear couple stress model for periodic sandwich
  beams}.
\newblock \bibinfo{journal}{Compos. Struct.} \bibinfo{volume}{212},
  \bibinfo{pages}{586--597}.
\bibitem[{Goncalves and Romanoff(2018)}]{goncalves2018}
\bibinfo{author}{Goncalves, B.R.}, \bibinfo{author}{Romanoff, J.},
  \bibinfo{year}{2018}.
\newblock \bibinfo{title}{Size-dependent modelling of elastic sandwich beams
  with prismatic cores}.
\newblock \bibinfo{journal}{Int. J. Solids Struct.} \bibinfo{volume}{136},
  \bibinfo{pages}{28--37}.
\bibitem[{Greer and Deshpande(2019)}]{greer2019}
\bibinfo{author}{Greer, J.}, \bibinfo{author}{Deshpande, V.},
  \bibinfo{year}{2019}.
\newblock \bibinfo{title}{Three-dimensional architected materials and
  structures: {D}esign, fabrication and mechanical behavior}.
\newblock \bibinfo{journal}{MRS Bull} \bibinfo{volume}{44},
  \bibinfo{pages}{750--757}.
\bibitem[{Hefzy and Nayfeh(1986)}]{hefzy1986}
\bibinfo{author}{Hefzy, M.S.}, \bibinfo{author}{Nayfeh, A.H.},
  \bibinfo{year}{1986}.
\newblock \bibinfo{title}{Shear deformation plate continua of large double
  layered space structures}.
\newblock \bibinfo{journal}{Int. J. Solids Struct.} \bibinfo{volume}{22},
  \bibinfo{pages}{1455--1469}.
\bibitem[{Kanatani(1979)}]{kanatani1979}
\bibinfo{author}{Kanatani, K.}, \bibinfo{year}{1979}.
\newblock \bibinfo{title}{A micropolar continuum model for vibrating grid
  frameworks}.
\newblock \bibinfo{journal}{Int. J. Eng. Sci.} \bibinfo{volume}{17},
  \bibinfo{pages}{409--418}.
\bibitem[{Karttunen et~al.(2017)Karttunen, Kanerva, Frank, Romanoff, Remes,
  Jelovica, Bossuyt and Sarlin}]{karttunen2017b}
\bibinfo{author}{Karttunen, A.T.}, \bibinfo{author}{Kanerva, M.},
  \bibinfo{author}{Frank, D.}, \bibinfo{author}{Romanoff, J.},
  \bibinfo{author}{Remes, H.}, \bibinfo{author}{Jelovica, J.},
  \bibinfo{author}{Bossuyt, S.}, \bibinfo{author}{Sarlin, E.},
  \bibinfo{year}{2017}.
\newblock \bibinfo{title}{Fatigue strength of laser-welded foam-filled steel
  sandwich beams}.
\newblock \bibinfo{journal}{Mater. Design.,} \bibinfo{volume}{115},
  \bibinfo{pages}{64--72}.
\bibitem[{Karttunen et~al.(2019a)Karttunen, Reddy and
  Romanoff}]{karttunen2019a}
\bibinfo{author}{Karttunen, A.T.}, \bibinfo{author}{Reddy, J.N.},
  \bibinfo{author}{Romanoff, J.}, \bibinfo{year}{2019}a.
\newblock \bibinfo{title}{Two-scale constitutive modeling of a lattice core
  sandwich beam}.
\newblock \bibinfo{journal}{Compos. B-Eng.} \bibinfo{volume}{160},
  \bibinfo{pages}{66--75}.
\bibitem[{Karttunen et~al.(2019b)Karttunen, Reddy and
  Romanoff}]{karttunen2019b}
\bibinfo{author}{Karttunen, A.T.}, \bibinfo{author}{Reddy, J.N.},
  \bibinfo{author}{Romanoff, J.}, \bibinfo{year}{2019}b.
\newblock \bibinfo{title}{Two-scale micropolar plate model for web-core
  sandwich panels}.
\newblock \bibinfo{journal}{Int. J. Solids Struct.} \bibinfo{volume}{170},
  \bibinfo{pages}{82--94}.
\bibitem[{Khakalo and Niiranen(2019)}]{khakalo2019}
\bibinfo{author}{Khakalo, S.}, \bibinfo{author}{Niiranen, J.},
  \bibinfo{year}{2019}.
\newblock \bibinfo{title}{Lattice structures as thermoelastic strain gradient
  metamaterials: {E}vidence from full-field simulations and applications to
  functionally step-wise-graded beams}.
\newblock \bibinfo{journal}{Compos. Part B-Eng.} \bibinfo{volume}{177, 107224}.
\bibitem[{Khakalo and Niiranen(2020)}]{khakalo2020}
\bibinfo{author}{Khakalo, S.}, \bibinfo{author}{Niiranen, J.},
  \bibinfo{year}{2020}.
\newblock \bibinfo{title}{Anisotropic strain gradient thermoelasticity for
  cellular structures: plate models, homogenization and isogeometric analysis}.
\newblock \bibinfo{journal}{J. Mech. Phys. Solids} \bibinfo{volume}{134,
  103728}.
\bibitem[{Kim and Piziali(1987)}]{kim1987}
\bibinfo{author}{Kim, K.S.}, \bibinfo{author}{Piziali, R.L.},
  \bibinfo{year}{1987}.
\newblock \bibinfo{title}{Continuum models of materials with
  beam-microstructure}.
\newblock \bibinfo{journal}{Int. J. Solids Struct.} \bibinfo{volume}{23},
  \bibinfo{pages}{1563--1578}.
\bibitem[{Kouznetsova et~al.(2002)Kouznetsova, Geers and
  Brekelmans}]{kouznetsova2002}
\bibinfo{author}{Kouznetsova, V.}, \bibinfo{author}{Geers, M.G.D.},
  \bibinfo{author}{Brekelmans, W.A.M.}, \bibinfo{year}{2002}.
\newblock \bibinfo{title}{Multi-scale constitutive modelling of heterogeneous
  materials with a gradient-enhanced computational homogenization scheme}.
\newblock \bibinfo{journal}{Int. J. Numer. Meth. Eng.} \bibinfo{volume}{54},
  \bibinfo{pages}{1235--1260}.
\bibitem[{Kumar and McDowell(2004)}]{kumar2004}
\bibinfo{author}{Kumar, R.S.}, \bibinfo{author}{McDowell, D.L.},
  \bibinfo{year}{2004}.
\newblock \bibinfo{title}{Generalized continuum modeling of 2-{D} periodic
  cellular solids}.
\newblock \bibinfo{journal}{Int. J. Solids Struct.} \bibinfo{volume}{41},
  \bibinfo{pages}{7399--7422}.
\bibitem[{Larsson and Diebels(2007)}]{larsson2007}
\bibinfo{author}{Larsson, R.}, \bibinfo{author}{Diebels, S.},
  \bibinfo{year}{2007}.
\newblock \bibinfo{title}{A second-order homogenization procedure for
  multi-scale analysis based on micropolar kinematics}.
\newblock \bibinfo{journal}{Int. J. Numer. Meth. Eng.} \bibinfo{volume}{69},
  \bibinfo{pages}{2485--2512}.
\bibitem[{Liu et~al.(2020)Liu, Wang, Jin, Liu and Lu}]{liu2020}
\bibinfo{author}{Liu, F.}, \bibinfo{author}{Wang, L.}, \bibinfo{author}{Jin,
  D.}, \bibinfo{author}{Liu, X.}, \bibinfo{author}{Lu, P.},
  \bibinfo{year}{2020}.
\newblock \bibinfo{title}{Equivalent micropolar beam model for spatial
  vibration analysis of planar repetitive truss structure with flexible
  joints}.
\newblock \bibinfo{journal}{Int. J. Mech. Sci.} \bibinfo{volume}{165, 105202}.
\bibitem[{Liu et~al.(2017)Liu, Kamm, Garc{\'\i}a-Moreno, Banhart and
  Pasini}]{liu2017}
\bibinfo{author}{Liu, L.}, \bibinfo{author}{Kamm, P.},
  \bibinfo{author}{Garc{\'\i}a-Moreno, F.}, \bibinfo{author}{Banhart, J.},
  \bibinfo{author}{Pasini, D.}, \bibinfo{year}{2017}.
\newblock \bibinfo{title}{Elastic and failure response of imperfect
  three-dimensional metallic lattices: the role of geometric defects induced by
  selective laser melting}.
\newblock \bibinfo{journal}{J. Mech. Phys. Solids} \bibinfo{volume}{107},
  \bibinfo{pages}{160--184}.
\bibitem[{Lu and Cary(2000)}]{lu2000}
\bibinfo{author}{Lu, H.}, \bibinfo{author}{Cary, P.D.}, \bibinfo{year}{2000}.
\newblock \bibinfo{title}{Deformation measurements by digital image
  correlation: implementation of a second-order displacement gradient}.
\newblock \bibinfo{journal}{Exp. Mech.} \bibinfo{volume}{40},
  \bibinfo{pages}{393--400}.
\bibitem[{Matou{\v{s}} et~al.(2017)Matou{\v{s}}, Geers, Kouznetsova and
  Gillman}]{matouvs2017}
\bibinfo{author}{Matou{\v{s}}, K.}, \bibinfo{author}{Geers, M.G.D.},
  \bibinfo{author}{Kouznetsova, V.G.}, \bibinfo{author}{Gillman, A.},
  \bibinfo{year}{2017}.
\newblock \bibinfo{title}{A review of predictive nonlinear theories for
  multiscale modeling of heterogeneous materials}.
\newblock \bibinfo{journal}{J. Comput. Phys.} \bibinfo{volume}{330},
  \bibinfo{pages}{192--220}.
\bibitem[{Mindlin(1963)}]{mindlin1963}
\bibinfo{author}{Mindlin, R.D.}, \bibinfo{year}{1963}.
\newblock \bibinfo{title}{Influence of couple-stresses on stress
  concentrations}.
\newblock \bibinfo{journal}{Exp. Mech.} \bibinfo{volume}{3},
  \bibinfo{pages}{1--7}.
\bibitem[{Mindlin and Tiersten(1962)}]{mindlin1962}
\bibinfo{author}{Mindlin, R.D.}, \bibinfo{author}{Tiersten, H.F.},
  \bibinfo{year}{1962}.
\newblock \bibinfo{title}{Effects of couple-stresses in linear elasticity}.
\newblock \bibinfo{journal}{Arch. Rational Mech. Anal.} \bibinfo{volume}{11},
  \bibinfo{pages}{415--448}.
\bibitem[{Mines(2019)}]{mines2019}
\bibinfo{author}{Mines, R.}, \bibinfo{year}{2019}.
\newblock \bibinfo{title}{Metallic Microlattice Structures}.
\newblock \bibinfo{publisher}{Springer}.
\bibitem[{Monforton and Wu(1963)}]{monforton1963}
\bibinfo{author}{Monforton, G.R.}, \bibinfo{author}{Wu, T.H.},
  \bibinfo{year}{1963}.
\newblock \bibinfo{title}{Matrix analysis of semi-rigid connected frames}.
\newblock \bibinfo{journal}{J. Struct. Div-ASCE} \bibinfo{volume}{89},
  \bibinfo{pages}{13--24}.
\bibitem[{Nampally et~al.(2019)Nampally, Karttunen and Reddy}]{nampally2019}
\bibinfo{author}{Nampally, P.}, \bibinfo{author}{Karttunen, A.T.},
  \bibinfo{author}{Reddy, J.N.}, \bibinfo{year}{2019}.
\newblock \bibinfo{title}{Nonlinear finite element analysis of lattice core
  sandwich beams}.
\newblock \bibinfo{journal}{Eur. J. Mech A-Solid.} \bibinfo{volume}{74},
  \bibinfo{pages}{431--439}.
\bibitem[{Nampally et~al.(2020)Nampally, Karttunen and Reddy}]{nampally2020}
\bibinfo{author}{Nampally, P.}, \bibinfo{author}{Karttunen, A.T.},
  \bibinfo{author}{Reddy, J.N.}, \bibinfo{year}{2020}.
\newblock \bibinfo{title}{Nonlinear finite element analysis of lattice core
  sandwich plates}.
\newblock \bibinfo{journal}{Int. J. Nonlin. Mech.} \bibinfo{volume}{121,
  103423}.
\bibitem[{Nilsson et~al.(2017)Nilsson, Al-Emrani and Atashipour}]{nilsson2017}
\bibinfo{author}{Nilsson, P.}, \bibinfo{author}{Al-Emrani, M.},
  \bibinfo{author}{Atashipour, S.R.}, \bibinfo{year}{2017}.
\newblock \bibinfo{title}{Transverse shear stiffness of corrugated core steel
  sandwich panels with dual weld lines}.
\newblock \bibinfo{journal}{Thin Wall. Struct.,} \bibinfo{volume}{117},
  \bibinfo{pages}{98--112}.
\bibitem[{Nilsson et~al.(2020)Nilsson, Al-Emrani and Atashipour}]{nilsson2020}
\bibinfo{author}{Nilsson, P.}, \bibinfo{author}{Al-Emrani, M.},
  \bibinfo{author}{Atashipour, S.R.}, \bibinfo{year}{2020}.
\newblock \bibinfo{title}{Fatigue-strength assessment of laser welds in
  corrugated core steel sandwich panels}.
\newblock \bibinfo{journal}{Journal of Constructional Steel Research}
  \bibinfo{volume}{164, 105797}.
\bibitem[{Niu and Yan(2016)}]{niu2016}
\bibinfo{author}{Niu, B.}, \bibinfo{author}{Yan, J.}, \bibinfo{year}{2016}.
\newblock \bibinfo{title}{A new micromechanical approach of micropolar
  continuum modeling for 2-{D} periodic cellular material}.
\newblock \bibinfo{journal}{Acta Mech Sinica} \bibinfo{volume}{32},
  \bibinfo{pages}{456--468}.
\bibitem[{Noor(1988)}]{noor1988}
\bibinfo{author}{Noor, A.K.}, \bibinfo{year}{1988}.
\newblock \bibinfo{title}{Continuum modeling for repetitive lattice
  structures}.
\newblock \bibinfo{journal}{Appl. Mech. Rev.} \bibinfo{volume}{41},
  \bibinfo{pages}{285--296}.
\bibitem[{Noor and Nemeth(1980)}]{noor1980}
\bibinfo{author}{Noor, A.K.}, \bibinfo{author}{Nemeth, M.P.},
  \bibinfo{year}{1980}.
\newblock \bibinfo{title}{Micropolar beam models for lattice grids with rigid
  joints}.
\newblock \bibinfo{journal}{Comput. Meth. Appl. Mech. Eng.}
  \bibinfo{volume}{21}, \bibinfo{pages}{249--263}.
\bibitem[{Park and Gao(2008)}]{park2008}
\bibinfo{author}{Park, S.K.}, \bibinfo{author}{Gao, X.L.},
  \bibinfo{year}{2008}.
\newblock \bibinfo{title}{Micromechanical modeling of honeycomb structures
  based on a modified couple stress theory}.
\newblock \bibinfo{journal}{Mech. Adv. Mater. Struc.} \bibinfo{volume}{15},
  \bibinfo{pages}{574--593}.
\bibitem[{Pasini and Guest(2019)}]{pasini2019}
\bibinfo{author}{Pasini, D.}, \bibinfo{author}{Guest, J.K.},
  \bibinfo{year}{2019}.
\newblock \bibinfo{title}{Imperfect architected materials: {M}echanics and
  topology optimization}.
\newblock \bibinfo{journal}{MRS Bull} \bibinfo{volume}{44},
  \bibinfo{pages}{766--772}.
\bibitem[{Penta et~al.(2017)Penta, Monaco, Pucillo and Gesualdo}]{penta2017}
\bibinfo{author}{Penta, F.}, \bibinfo{author}{Monaco, M.},
  \bibinfo{author}{Pucillo, G.P.}, \bibinfo{author}{Gesualdo, A.},
  \bibinfo{year}{2017}.
\newblock \bibinfo{title}{Periodic beam-like structures homogenization by
  transfer matrix eigen-analysis: A direct approach}.
\newblock \bibinfo{journal}{Mech. Res. Commun.} \bibinfo{volume}{85},
  \bibinfo{pages}{81--88}.
\bibitem[{Phani and Hussein(2017)}]{phani2017}
\bibinfo{author}{Phani, A.S.}, \bibinfo{author}{Hussein, M.I.},
  \bibinfo{year}{2017}.
\newblock \bibinfo{title}{Dynamics of lattice materials}.
\newblock \bibinfo{publisher}{Wiley}.
\bibitem[{Rahali et~al.(2016)Rahali, Goda and Ganghoffer}]{rahali2016}
\bibinfo{author}{Rahali, Y.}, \bibinfo{author}{Goda, I.},
  \bibinfo{author}{Ganghoffer, J.F.}, \bibinfo{year}{2016}.
\newblock \bibinfo{title}{Numerical identification of classical and
  nonclassical moduli of 3{D} woven textiles and analysis of scale effects}.
\newblock \bibinfo{journal}{Compos. Struct} \bibinfo{volume}{135},
  \bibinfo{pages}{122--139}.
\bibitem[{Reda et~al.(2017)Reda, Ganghoffer and Lakiss}]{reda2017}
\bibinfo{author}{Reda, H.}, \bibinfo{author}{Ganghoffer, J.F.},
  \bibinfo{author}{Lakiss, H.}, \bibinfo{year}{2017}.
\newblock \bibinfo{title}{Micropolar dissipative models for the analysis of
  2{D} dispersive waves in periodic lattices}.
\newblock \bibinfo{journal}{J. Sound Vib} \bibinfo{volume}{392},
  \bibinfo{pages}{325--345}.
\bibitem[{Reddy(2011)}]{reddy2011}
\bibinfo{author}{Reddy, J.N.}, \bibinfo{year}{2011}.
\newblock \bibinfo{title}{Microstructure-dependent couple stress theories of
  functionally graded beams}.
\newblock \bibinfo{journal}{J. Mech. Phys. Solids} \bibinfo{volume}{59},
  \bibinfo{pages}{2382--2399}.
\bibitem[{Reddy(2018)}]{reddy2018}
\bibinfo{author}{Reddy, J.N.}, \bibinfo{year}{2018}.
\newblock \bibinfo{title}{{Energy Principles and Variational Methods in Applied
  Mechanics}}.
\newblock \bibinfo{edition}{3rd} ed., \bibinfo{publisher}{John Wiley\& Sons},
  \bibinfo{address}{New York, NY}.
\bibitem[{Roland and Metschkow(1997)}]{roland1997}
\bibinfo{author}{Roland, F.}, \bibinfo{author}{Metschkow, B.},
  \bibinfo{year}{1997}.
\newblock \bibinfo{title}{Laser welded sandwich panels for shipbuilding and
  structural steel engineering}.
\newblock \bibinfo{publisher}{Transactions on the Built Environment, vol 24.
  WIT Press}.
\bibitem[{Romanoff and Reddy(2014)}]{romanoff2014}
\bibinfo{author}{Romanoff, J.}, \bibinfo{author}{Reddy, J.N.},
  \bibinfo{year}{2014}.
\newblock \bibinfo{title}{Experimental validation of the modified couple stress
  {T}imoshenko beam theory for web-core sandwich panels}.
\newblock \bibinfo{journal}{Compos. Struct.} \bibinfo{volume}{111},
  \bibinfo{pages}{130--137}.
\bibitem[{Romanoff et~al.(2007)Romanoff, Remes, Socha, Jutila and
  Varsta}]{romanoff2007c}
\bibinfo{author}{Romanoff, J.}, \bibinfo{author}{Remes, H.},
  \bibinfo{author}{Socha, G.}, \bibinfo{author}{Jutila, M.},
  \bibinfo{author}{Varsta, P.}, \bibinfo{year}{2007}.
\newblock \bibinfo{title}{The stiffness of laser stake welded {T}-joints in
  web-core sandwich structures}.
\newblock \bibinfo{journal}{Thin Wall. Struct.} \bibinfo{volume}{45},
  \bibinfo{pages}{453--462}.
\bibitem[{Sadd(2014)}]{sadd2014}
\bibinfo{author}{Sadd, M.H.}, \bibinfo{year}{2014}.
\newblock \bibinfo{title}{{Elasticity -- Theory, Applications and Numerics}}.
\newblock \bibinfo{edition}{3rd} ed., \bibinfo{publisher}{Academic Press},
  \bibinfo{address}{Oxford}.
\bibitem[{Salehian and Inman(2010)}]{salehian2010}
\bibinfo{author}{Salehian, A.}, \bibinfo{author}{Inman, D.J.},
  \bibinfo{year}{2010}.
\newblock \bibinfo{title}{Micropolar continuous modeling and frequency response
  validation of a lattice structure}.
\newblock \bibinfo{journal}{J. Vib. Acoust.} \bibinfo{volume}{132}.
\bibitem[{Schaedler and Carter(2016)}]{carter2016}
\bibinfo{author}{Schaedler, T.A.}, \bibinfo{author}{Carter, W.B.},
  \bibinfo{year}{2016}.
\newblock \bibinfo{title}{Architected cellular materials}.
\newblock \bibinfo{journal}{Ann. Rev. Mater. Res.} \bibinfo{volume}{46},
  \bibinfo{pages}{187--210}.
\bibitem[{Shaat and Abdelkefi(2016)}]{shaat2016}
\bibinfo{author}{Shaat, M.}, \bibinfo{author}{Abdelkefi, A.},
  \bibinfo{year}{2016}.
\newblock \bibinfo{title}{On a second-order rotation gradient theory for linear
  elastic continua}.
\newblock \bibinfo{journal}{Int. J. Eng. Sci.} \bibinfo{volume}{100},
  \bibinfo{pages}{74--98}.
\bibitem[{Spadoni and Ruzzene(2012)}]{spadoni2012}
\bibinfo{author}{Spadoni, A.}, \bibinfo{author}{Ruzzene, M.},
  \bibinfo{year}{2012}.
\newblock \bibinfo{title}{Elasto-static micropolar behavior of a chiral auxetic
  lattice}.
\newblock \bibinfo{journal}{J. Mech. Phys. Solids} \bibinfo{volume}{60},
  \bibinfo{pages}{156--171}.
\bibitem[{St-Pierre et~al.(2015)St-Pierre, Fleck and Deshpande}]{luc2015}
\bibinfo{author}{St-Pierre, L.}, \bibinfo{author}{Fleck, N.A.},
  \bibinfo{author}{Deshpande, V.S.}, \bibinfo{year}{2015}.
\newblock \bibinfo{title}{The dynamic indentation response of sandwich panels
  with a corrugated or {Y}-frame core}.
\newblock \bibinfo{journal}{Int. J. Mech. Sci.} \bibinfo{volume}{92},
  \bibinfo{pages}{279--289}.
\bibitem[{Sun and Yang(1973)}]{sun1973}
\bibinfo{author}{Sun, C.T.}, \bibinfo{author}{Yang, T.Y.},
  \bibinfo{year}{1973}.
\newblock \bibinfo{title}{A continuum approach toward dynamics of gridworks}.
\newblock \bibinfo{journal}{J. Appl. Mech.} \bibinfo{volume}{40},
  \bibinfo{pages}{186--192}.
\bibitem[{Sun and Yang(1975)}]{sun1975}
\bibinfo{author}{Sun, C.T.}, \bibinfo{author}{Yang, T.Y.},
  \bibinfo{year}{1975}.
\newblock \bibinfo{title}{A couple-stress theory for gridwork-reinforced
  media}.
\newblock \bibinfo{journal}{J. Elasticity} \bibinfo{volume}{5},
  \bibinfo{pages}{45--58}.
\bibitem[{Teasdale(1988)}]{teasdale1988}
\bibinfo{author}{Teasdale, J.A.}, \bibinfo{year}{1988}.
\newblock \bibinfo{title}{The application of sandwich structures to ship
  design: phase four summary report}.
\newblock \bibinfo{type}{Technical Report}. Department of Naval Architecture
  and Shipbuilding, University of Newcastle-upon-Tyne.
\bibitem[{Tiersten and Bleustein(1974)}]{tiersten1974}
\bibinfo{author}{Tiersten, H.F.}, \bibinfo{author}{Bleustein, J.L.},
  \bibinfo{year}{1974}.
\newblock \bibinfo{title}{Generalized Elastic Continua, In: Herrmann G. (Ed.),
  R. D. Mindlin and Applied Mechanics}.
\newblock \bibinfo{publisher}{Pergamon Press}, \bibinfo{address}{New York}.
\bibitem[{Toupin(1962)}]{toupin1962}
\bibinfo{author}{Toupin, R.A.}, \bibinfo{year}{1962}.
\newblock \bibinfo{title}{Elastic materials with couple-stresses}.
\newblock \bibinfo{journal}{Arch. Rational Mech. Anal.} \bibinfo{volume}{11},
  \bibinfo{pages}{385--414}.
\bibitem[{Valdevit et~al.(2018)Valdevit, Bertoldi, Guest and
  Spadaccini}]{valdevit2018}
\bibinfo{author}{Valdevit, L.}, \bibinfo{author}{Bertoldi, K.},
  \bibinfo{author}{Guest, J.}, \bibinfo{author}{Spadaccini, C.},
  \bibinfo{year}{2018}.
\newblock \bibinfo{title}{Architected materials: {S}ynthesis, characterization,
  modeling and optimal design}.
\newblock \bibinfo{journal}{J. Mater. Res.} \bibinfo{volume}{33},
  \bibinfo{pages}{241--246}.
\bibitem[{Wadley(2006)}]{wadley2006}
\bibinfo{author}{Wadley, H.}, \bibinfo{year}{2006}.
\newblock \bibinfo{title}{Multifunctional periodic cellular metals}.
\newblock \bibinfo{journal}{Philos. T. R. Soc. A.} \bibinfo{volume}{364},
  \bibinfo{pages}{31--68}.
\bibitem[{Yang et~al.(2002)Yang, Chong, Lam and Tong}]{yang2002}
\bibinfo{author}{Yang, F.}, \bibinfo{author}{Chong, A.C.M.},
  \bibinfo{author}{Lam, D.C.C.}, \bibinfo{author}{Tong, P.},
  \bibinfo{year}{2002}.
\newblock \bibinfo{title}{Couple stress based strain gradient theory for
  elasticity}.
\newblock \bibinfo{journal}{Int. J. Solids Struct.} \bibinfo{volume}{39},
  \bibinfo{pages}{2731--2743}.
\bibitem[{Zenkert(1997)}]{zenkert1997}
\bibinfo{author}{Zenkert, D.}, \bibinfo{year}{1997}.
\newblock \bibinfo{title}{The handbook of sandwich construction}.
\newblock \bibinfo{publisher}{Engineering Materials Advisory Services}.
\bibitem[{Zhu(2010)}]{zhu2010}
\bibinfo{author}{Zhu, H.X.}, \bibinfo{year}{2010}.
\newblock \bibinfo{title}{Size-dependent elastic properties of micro-and
  nano-honeycombs}.
\newblock \bibinfo{journal}{J. Mech. Phys. Solids} \bibinfo{volume}{58},
  \bibinfo{pages}{696--709}.

\end{thebibliography}





\end{document}